\documentclass[usenatbib]{mn2e}
\usepackage{epsfig}
\usepackage{amsmath}

\def\gsim{\mathrel{\raise0.35ex\hbox{$\scriptstyle >$}\kern-0.6em 
\lower0.40ex\hbox{{$\scriptstyle \sim$}}}}
\def\lsim{\mathrel{\raise0.35ex\hbox{$\scriptstyle <$}\kern-0.6em 
\lower0.40ex\hbox{{$\scriptstyle \sim$}}}}
\def\gs{\mathrel{\raise0.35ex\hbox{$\scriptstyle >$}\kern-0.6em 
\lower0.40ex\hbox{{$\scriptstyle \sim$}}}}
\def\ls{\mathrel{\raise0.35ex\hbox{$\scriptstyle <$}\kern-0.6em 
\lower0.40ex\hbox{{$\scriptstyle \sim$}}}}

\def\kms {{\,\rm km\,s^{-1}}} 

\def\lesssim{\mathrel{\hbox{\rlap{\hbox{\lower4pt\hbox{$\sim$}}}\hbox{$<$}}}}
\def\gtrsim{\mathrel{\hbox{\rlap{\hbox{\lower4pt\hbox{$\sim$}}}\hbox{$>$}}}}

\date{\today}
\title[The Build-up of the Colour-Magnitude Relation]
{The Build-up of the Colour-Magnitude Relation as a Function of Environment}

\author[Tanaka et al.]{
\parbox[t]{\textwidth}{
Masayuki Tanaka$^1$,
Tadayuki Kodama$^{2,3}$,
Nobuo Arimoto$^2$,
Sadanori Okamura$^{1,4}$,
Keiichi Umetsu$^5$,
Kazuhiro Shimasaku$^{1,4}$,
Ichi Tanaka$^6$,
Toru Yamada$^2$
}
\vspace*{6pt}\\
$^{1}$Department of Astronomy, School of Science, University of Tokyo, Tokyo 113--0033, Japan \\
$^{2}$National Astronomical Observatory of Japan, Mitaka, Tokyo 181--8588, Japan \\
$^{3}$European Southern Observatory, Karl-Schwarzschild-Str. 2, D-85748, Garching, Germany\\
$^{4}$Research Center for the Early Universe, School of Science, University of Tokyo, Tokyo 113--0033, Japan\\
$^{5}$Institute of Astronomy and Astrophysics, Academia Sinica, Taipei 106, Taiwan\\
$^{6}$Astronomical Institute, Tohoku University, Aoba-ku, Sendai 980--8578, Japan\\
}

\begin{document}

\maketitle

\begin{abstract}
We discuss the environmental dependence of galaxy evolution based on
deep panoramic imaging of two distant clusters,
RXJ0152.7$-$1257 at $z=0.83$ and CL0016+1609 at $z=0.55$, taken with
Suprime-Cam on the Subaru Telescope as part of the {\it PISCES} project.
By combining with the {\it SDSS} data as a local counterpart for
comparison, we construct a large sample of galaxies that spans wide
ranges in environment, time, and stellar mass (or luminosity).  This
allows us to conduct systematic and statistical analyses of the
photometric properties of galaxies based on the colour--density
diagrams, colour--magnitude relations, and luminosity functions.
We find that colours of galaxies, especially those of faint
galaxies ($M_V>M_V^*+1$), change from blue to red at a
break density as we go to denser regions.  This trend is observed
at all redshifts in our sample.  Based on local and global densities
of galaxies, we classify three environments: 
field, groups, and clusters, and look into the
environmental dependence of galaxies in detail.  In particular, we
quantify how the colour-magnitude relation is built-up as
a function of environment.  We show that the bright-end of the {\it cluster}
colour-magnitude relation is already built at $z=0.83$,
while the faint-end is possibly still in the process of build-up.
In contrast to this, the bright-end of the {\it field} colour-magnitude relation
has been vigorously built all the way down to the present-day
and the build-up at the faint-end has not started yet.
A possible interpretation of these results is that galaxies evolve
in the 'down-sizing' fashion.
That is, massive galaxies complete their star formation first and the
truncation of star formation is propagated to smaller objects as time progresses.
This trend is likely to depend on environment since the build-up of the
colour-magnitude relation is delayed in lower-density environments.
Therefore, we may suggest that the evolution of galaxies took place earliest
in massive galaxies and in high density regions, and it is delayed in
less massive galaxies and in lower density regions.
Further studies are, however, obviously needed to confirm the observed trends
and establish the 'down-sizing' picture.
\end{abstract}


%
%
\section{Introduction}

\label{sec:intro}
Intensive studies on galaxy properties, such as star formation rates
and morphology,  have significantly improved 
our understanding of galaxies in the Universe.
It is, however, still unclear how galaxies evolve over the Hubble time.
This is due to the complex nature of galaxy properties;
galaxy properties depend not only on time, but also on environment and mass (luminosity).
These three axes are related to one another
and they characterize galaxy evolution.
Despite the obvious importance, however, galaxy properties
along these three axes have not been viewed simultaneously and systematically,
due to the limited previous data sets none of which can cover all of these three axes.
Based on the wide-field Subaru data and the large SDSS data,
we present in this paper a comprehensive study of star formation
activity of galaxies along the three axes.

It is now well known that galaxy properties depend on environment in which
galaxies reside.
This environmental dependence of galaxy properties was first quantitatively
studied by \citet{dressler80}, who showed the morphology-density relation
based on 55 nearby clusters.
There is a clear trend that early-type galaxies are preferentially
located in high density regions, while late-type galaxies tend
to be located in lower density regions.
Following this work, intensive studies on environmental dependence of galaxy properties
have been carried out, and strengthened or extended the Dressler's result
(e.g., \citealt{postman84,whitmore93,balogh97,dressler97,balogh98,hashimoto98,lubin98,
oke98,postman98,vandokkum98,balogh99,poggianti99,lubin00,couch01,kodama01a,postman01,
lewis02,lubin02,blanton03a,gomez03,goto03a,hogg03,treu03,balogh04a,balogh04b,smith04,tanaka04}).

Galaxy properties are also known to depend on mass (luminosity) of galaxies.
Recently conducted large surveys, such as 2dF \citep{colless03}
and SDSS \citep{york00}, revealed that galaxy properties show strong
bimodality in their distribution
\citep{strateva01,blanton03a,kauffmann03,balogh04a,kauffmann04,tanaka04}.
That is, there are two distinct populations: red early-type galaxies
and blue late-type galaxies.
This bimodality is found to be a strong function of mass of galaxies
in the sense that massive galaxies tend to be red early-type galaxies,
while less massive galaxies tend to be blue late-type galaxies
\citep{kauffmann03,balogh04a,baldry04,bell04,kauffmann04}

Galaxies evolve with time, of course, as a natural consequence of stellar
evolution.
It is now 20 years since \citet{butcher84} presented their
startling result that the fraction of blue galaxies in clusters increases
with look-back time or redshift.
This B--O effect seems to be confirmed by later studies (\citealt{couch87,
rakos95,ellingson01,kodama01b,margoniner01,nakata01,fairley02,goto03b,tran04};
but see also \citealt{fairley02,andreon04}).
This effect suggests that star formation activities in cluster galaxies
change with time, in the sense that the fraction of star forming galaxies
decreases with time.

These three axes, that is, environment, mass, and time, must be closely related
to one another.
For example, the morphology-density relation is found to evolve
(\citealt{dressler97,fasano00,treu03,smith04}, but see also \citealt{andreon98}).
\citet{tanaka04} showed that environmental dependence
of star formation activity and morphology of galaxies change as a
function of galaxy luminosity (mass).
To untie this complexity, we base our analyses on panoramic imaging data
of two high redshift clusters, CL0015.9+16 at $z=0.55$ and RXJ0152.7--13
at $z=0.83$, and the SDSS data.
The former data are obtained as part of our on-going Subaru distant
cluster project called PISCES.
Our data span wide ranges in environment, time, and stellar mass,
and give us an unique opportunity to investigate star formation activity
along all the three axes simultaneously for the first time.

The structure of this paper is as follows.
In \S \ref{sec:data}, we briefly review our project PISCES and our data.
Also we describe the data of the local Universe that we use.
Then we move on to the procedure we adopt to
eliminate foreground and background contamination in \S \ref{sec:field_sub}.
Before presenting main results, we summarize various photometric and environmental
quantities used in this paper in \S \ref{sec:parameters}.
We examine environmental dependence of galaxy star formation in \S \ref{sec:colour-density}.
Based on results obtained in \S \ref{sec:colour-density},
we re-define environments in \S \ref{sec:env_def}.
Colour-magnitude diagrams and luminosity functions in various environments
are shown in \S \ref{sec:colour-magnitude} and \ref{sec:luminosity_function},
respectively.
In \S \ref{sec:discussion}, we discuss the implications of our results
on the formation and evolution of galaxies.
Finally, a summary is given in \S \ref{sec:conclusion}.
Most of the important conclusions in this paper are drawn from the results 
presented in \S 5.1,  \S 7, and \S 9.
Readers interested only in our primary conclusions can go to
these sections directly after brief reading of \S 2-4.

Throughout this paper, we assume a flat Universe of
$\Omega_{\rm M}=0.3,\ \Omega_{\rm \Lambda}=0.7$ and $H_0=70\kms \rm Mpc^{-1}$.
We use the AB magnitude system for observed quantities
and the Vega-referred system for rest-frame ones.
We use the following abbreviation: CMD for colour-magnitude diagram,
CMR for colour-magnitude relation, and LF for luminosity function.

%
%
\section{DATA}
\label{sec:data}
\subsection{PISCES Project}
We are conducting a systematic study of cluster evolution based on
panoramic multi-band imaging with Suprime-Cam and optical multi-slit spectroscopy
with FOCAS on Subaru.
In this section, we briefly review the project
{\it Panoramic Imaging and Spectroscopy of Cluster Evolution with Subaru} (PISCES).
The reader should refer to \citet{kodama05} for further details.

The primary aims of this project are two-fold:
(1) to map out large scale structures around distant clusters
to trace the cluster assembly history and (2) to look into galaxy properties in detail
as a function of environment along the structures in order to directly identify
the environmental effects acting on galaxies during their assembly to higher density regions.
The unique feature of this project is its wide-field coverage by taking advantage of
the Suprime-Cam \citep{miyazaki02}, which provides a $34'\times27'$ field of view.
Therefore, the PISCES will provide an opportunity to link an evolutionary path
between galaxies in the local Universe and high redshift counterparts
over a wide range in environment.
Also, the depth of PISCES, reaching down to $M^*+4$ at $z\sim1$ with an 8-m telescope,
will shed light on the nature of faint galaxies at high redshifts, which probably show
quite different properties compared with massive galaxies, as seen in the local Universe
(e.g., \citealt{baldry04,kauffmann04}).
As part of this on-going project,
we observed CL0015.9+1609 and RXJ0152.7-1357 in September 2003.

The cluster CL0015.9+1609 (CL0016 for short) is one of the most extensively studied galaxy clusters,
and it has been observed in various wavelength ranges: X-ray
(e.g., \citealt{worrall03}), UV \citep{brown00}, optical (e.g., \citealt{dressler99}),
and submm (e.g., \citealt{zemcov03}).
\citet{koo81} suggested the existence of a large-scale structure around the
cluster based on the photographic photometry, and later,
the cluster was found to have companion clusters \citep{hughes95,connolly96,hughes98}.
A clear CMR is seen \citep{ellis97,dahlen04}, and the cluster is
known to have a very low blue fraction \citep{butcher84}.

The cluster RXJ0152.7--1357 (RXJ0153 for short) is one of the most
X-ray luminous distant ($z>0.7$) clusters known.
The cluster was discovered independently in the {\it ROSAT Deep Cluster Survey}
(RDCS; \citealt{rosati98}) and in the {\it Wide Angle ROSAT Pointed Survey}
(WARPS; \citealt{scharf97}; \citealt{ebeling00}).
Later, it was also detected in the {\it Serendipitous High-redshift Archival ROSAT Cluster}
(SHARC) survey \citep{romer00}.
Since these discoveries, the cluster has been the subject of {\it BeppoSAX, Chandra}
and {\it XMM-Newton} observations \citep{dellaceca00,maughan03,jones03}.
Observations of the Sunyaev-Zeldovich effect
\citep{joy01}, spectroscopic follow-up \citep{demarco04,homeier04,jorgensen04},
near-IR imaging \citep{ellis04}, 
and the weak lensing analysis \citep{huo04,umetsu04,jee04} have also been performed.
All these studies show that RXJ0153 is indeed a very massive cluster at $z=0.83$:
the bolometric X-ray luminosity of $>1\times10^{45}\rm\ ergs\ s^{-1}$,
and the total dynamical mass of $\sim1\times10^{15}\ \rm M_{\odot}$.
Based on the ACS weak lensing analysis, \citet{jee04} suggested that
the previous cluster mass estimates may be an over-estimation.

Below, we briefly summarize our observation of these two clusters.
Details of the observation and data reduction are described in \citet{kodama05}.

The observing conditions were excellent during the nights
with a typical seeing size of $0''.6$.
CL0016 was observed in $BVRi'z'$ and RXJ0153 in $VRi'z'$.
Exposure times and limiting magnitudes are shown in Table \ref{tab:obs_summary}.
Object detection is performed using {\it SExtractor} (v.2.3.2; \citealt{bertin96}).
Objects in CL0016 are $i'$-band selected and those in RXJ0153 are $z'$-band selected.
It should be noted that the $i'$-band for CL0016 and the $z'$-band for RXJ0153
correspond to almost the same rest-frame wavelength range, and therefore
we have little difference in selection effects between the two clusters.
We use MAG\_AUTO for total magnitudes and $2''$ diameter aperture magnitudes for colours.
Magnitude zero-points determined by the photometric standard stars
taken during the same observing run are found to be offset
($\lesssim 0.1$ mag.) with respect to stellar SEDs of \citet{gunn83}.
We shift the zero-points so as to match with Gunn \& Stryker stars.
Star-galaxy separation is performed on the basis of FWHM vs. total magnitude diagrams.

In what follows, we restrict ourselves to $i'<24.5$ galaxies in CL0016 and
$z'<25.0$ galaxies in RXJ0153.
These magnitude cuts ensure that we are not affected by incompleteness effects.

\begin{table}
\begin{center}
\begin{tabular}{lccc}\hline
Cluster & Filter & Exposure Time (min.) &  Limiting Magnitude\\
\hline
CL0016  & $B$    & 90                   &  26.9\\
        & $V$    & 96                   &  26.2\\
        & $R$    & 64                   &  26.0\\
        & $i'$   & 60                   &  25.9\\
        & $z'$   & 47.5                 &  24.6\\
RXJ0153 & $V$    & 120                  &  26.7\\
        & $R$    & 116                  &  26.5\\
        & $i'$   & 75                   &  26.1\\
        & $z'$   & 77                   &  25.0\\
\hline
\end{tabular}
\caption{
A list of exposure times and limiting magnitudes (AB system).
Limiting magnitudes are shown as a $5\sigma$ limit
measured in a $2''$ aperture.
}
\label{tab:obs_summary}
\end{center}
\end{table}

\subsection{SDSS}
As a local counterpart of the two high-$z$ samples, we use the data from the Sloan Digital
Sky Survey (SDSS; \citealt{york00,stoughton02,abazajian03,abazajian04}).
The SDSS observes one quarter of the sky both photometrically and spectroscopically.
The imaging survey is performed in five optical bands, $u,\ g,\ r,\ i,\ $ and $z$
\citep{fukugita96,gunn98,hogg01,smith02,pier03}.
The spectroscopic survey is done with a pair of double fiber-fed spectrographs
which covers $3800\rm\AA\ -\ 9200\AA$.
Each fiber subtends $3''$ on the sky.
Since a fiber cannot be placed closer than $55''$ to a nearby fiber due to mechanical
constraints, the tiling algorithm has been developed to reduce the number of
unobserved objects \citep{blanton03b}.
The overall completeness of the spectroscopic survey is expected to be over 90\%.

We use the the public data of the second data release (DR2; \citealt{abazajian04}).
Galaxies in the Main Galaxy Sample \citep{strauss02} are used here.
The sample covers 2627 square degrees of the sky.
We extract galaxies at $0.005<z<0.065$.
We perform a volume correction to a flux limited sample taking into account
large-scale structures, instead of making a volume-limited sample, 
so that we can statistically reach as deep as the high-$z$ samples
in terms of magnitude relative to the characteristic magnitude.
Details of the volume correction is described in Appendix \ref{app:vmax}.
We use Petrosian magnitudes \citep{petrosian76,stoughton02} for total magnitudes
and model magnitudes for colours.
All these quantities are Galactic extinction corrected \citep{schlegel98},
and k-corrected using the code of 
Blanton et al. (2003c, v3\_2).

Main galaxies in the SDSS are $r$-band selected \citep{strauss02},
while those in RXJ0153 and CL0016 are selected in the rest-frame $\simeq g$-band.
To minimize selection effects, we apply the magnitude cut of $g<18$
to our SDSS sample so that the sample mimics a $g$-selected one
at the cost of reducing the number of sample galaxies.
This cut leaves 41695 galaxies.
In summary, the galaxies in RXJ0153, CL0016 and SDSS are all selected
at similar wavelengths in the rest-frame nearly corresponding to the $g$-band.

%
%
\section{Contamination Subtraction}
\label{sec:field_sub}

Since our two clusters from PISCES lie at high redshifts
and our imaging is deep,
galaxies at the cluster redshifts are heavily contaminated by
foreground and background galaxies.
Thus, in order to study galaxies at the cluster redshifts,
it is essential to eliminate the contamination.
Our strategy for the contamination subtraction is two-fold.
Firstly, we apply photometric redshift technique to largely eliminate
fore-/background galaxies (e.g., \citealt{kodama01a,nakata04}).
Although photometric redshift is a powerful tool,
the contamination remains at a non-negligible level.
We therefore statistically subtract the remaining contamination
on the basis of CMDs \citep{kodama01a,pimbblet02}.
Each procedure is described in the following subsections.

\subsection{Photometric Redshifts}
We apply the photometric redshift code of \citet{kodama99} to
the photometric catalogue of the PISCES clusters \citep{kodama05}.
The photometric redshift utilizes the population synthesis model of
\citet{kodama97}.
Star formation histories of model templates are described by a combination
of the elliptical galaxy model with a very short time scale (0.1Gyr)
of star formation and the disk model with a much longer time scale of
$\tau=5$ Gyr (star formation rate $\propto e^{-t/\tau}$).
The models are constructed so as to reproduce the observed colours of 
galaxies \citep{kodama99}.
It is found that the observed colours of the CMR are slightly offset
compared with the model colours.
We shifted the model zero-points systematically
so that the model colours of the CMR match with the observed
colours at the cluster redshifts.
These shifts are required to calibrate and improve our photometric redshifts.

\begin{figure*}
\begin{center}
\leavevmode
\epsfxsize 0.45\hsize \epsfbox{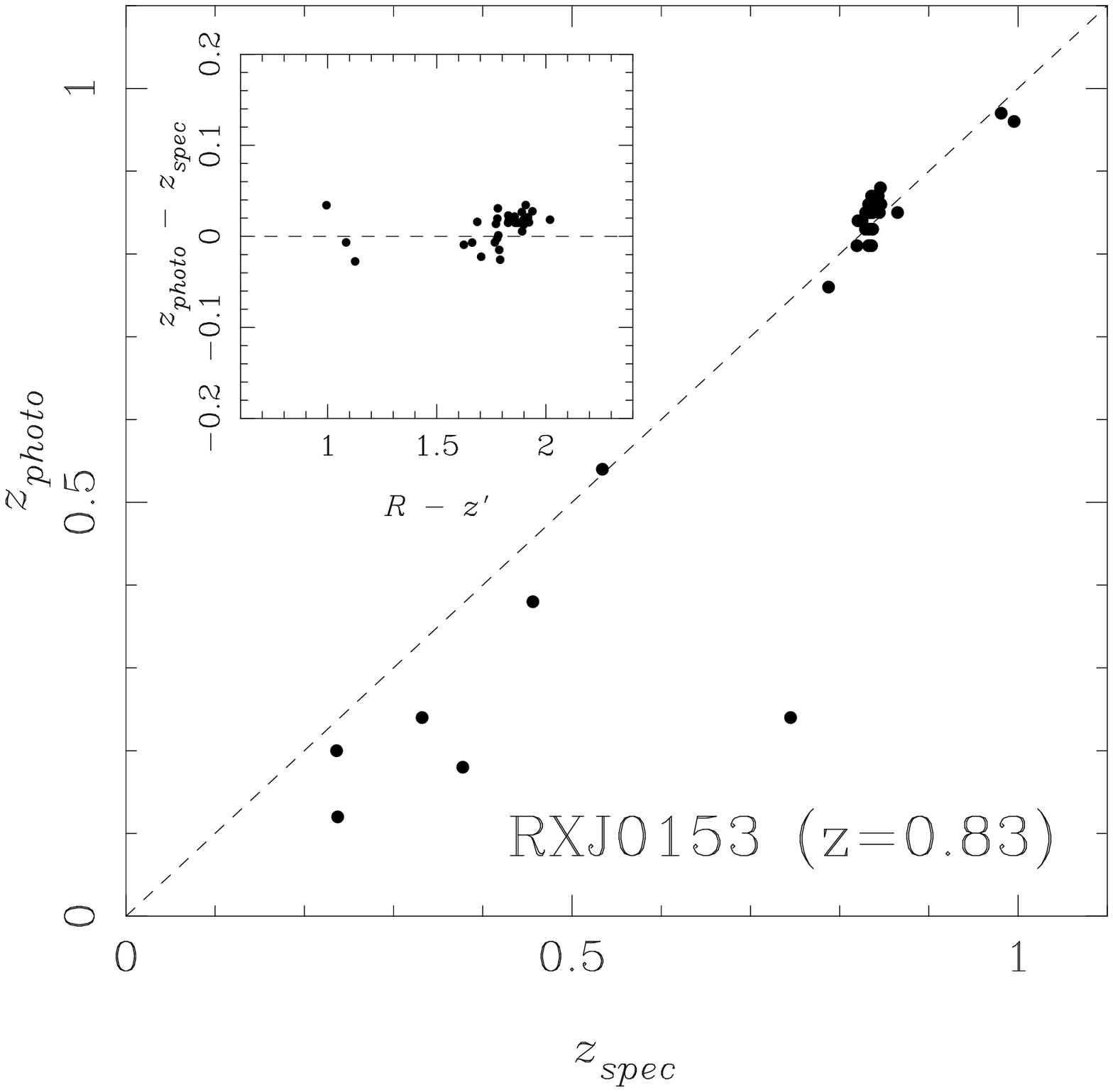}\hspace{0.5cm}
\epsfxsize 0.45\hsize \epsfbox{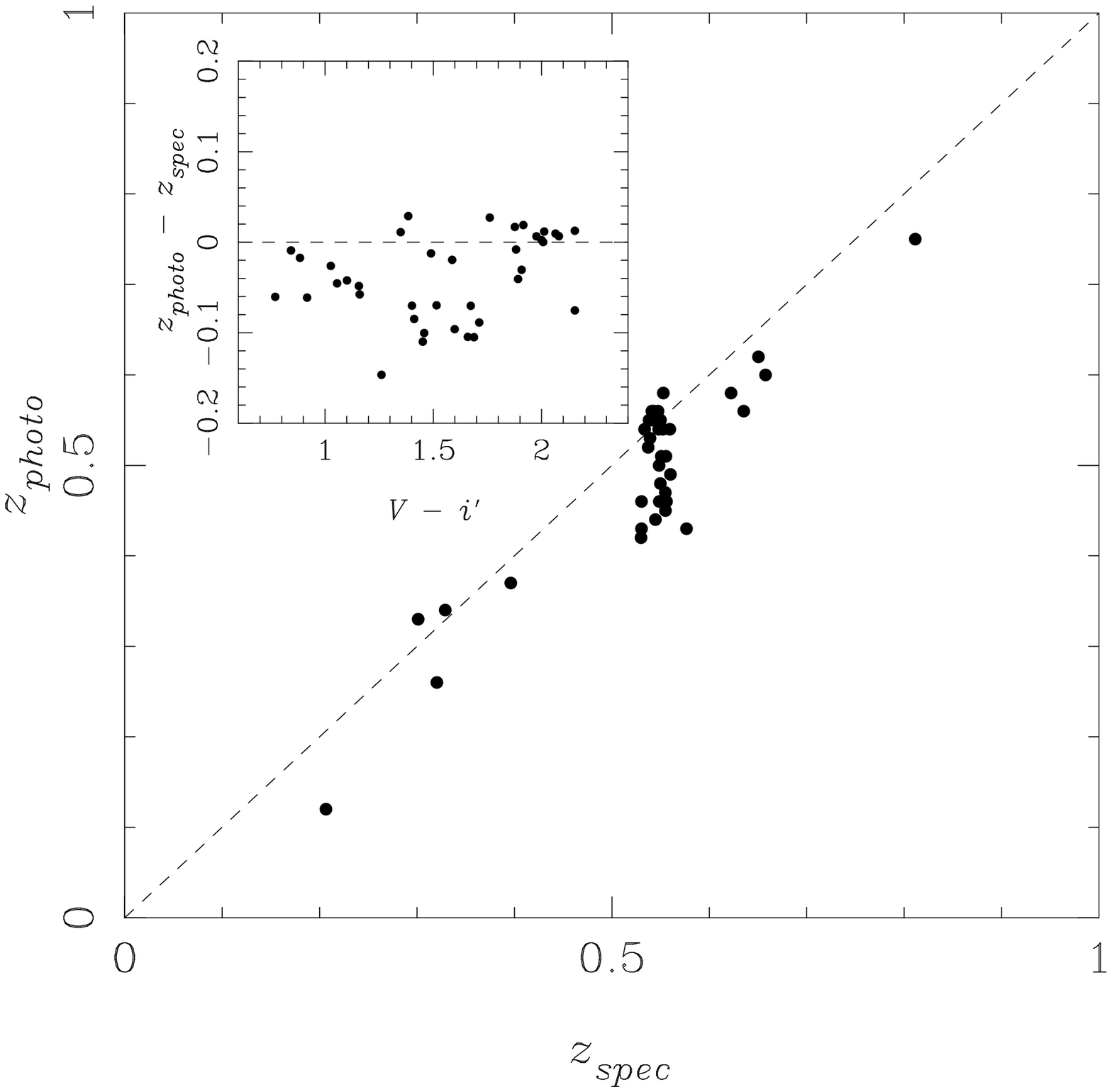}\\
\end{center}
\caption{
{\it Left:}
Photometric redshifts plotted against spectrally determined redshifts for RXJ0153.
In the inset, we show differences between photometric redshifts and
spectroscopic redshifts as a function of $R-z'$ colour, which is close to the rest-frame
$U-V$ colour, using galaxies at $0.78<z_{spec}<0.88$.
{\it Right:} Same as the left panel but for CL0016.
In the inset, we use $V-i'$ colour and galaxies at $0.5<z_{spec}<0.6$.
}
\label{fig:photoz_vs_specz}
\end{figure*}

To assess the accuracy of our photometric redshifts,
we compare the photometric redshifts with spectroscopically determined redshifts
as shown in Figure \ref{fig:photoz_vs_specz}.
For RXJ0153, we use the spectroscopic sample of \citet{jorgensen04}.
Our photometric redshifts are fairly good for cluster galaxies.
But, at $z<0.5$, the accuracy is rather poor since we lack $U$ and $B$-band data.
Excluding the most deviant galaxy at $z_{spec}=0.745$, the mean and median of
$z_{photo}-z_{spec}$ are $+0.0067$ and $+0.0093$, respectively.
The standard deviations around the mean and median are found to be $0.045$ and $0.047$.
For CL0016, we compile spectroscopic redshift data of \citet{hughes95},
\citet{munn97}, \citet{hughes98}, and \citet{dressler99}.
Excluding the two most deviant galaxies at $z_{spec}\sim0.4$, the mean and median of
$z_{photo}-z_{spec}$ are $+0.0036$ and $-0.0118$, respectively.
The standard deviations around the mean and median are $0.055$ and $0.057$.
The figures demonstrate that our photometric redshifts are fairly accurate.
We note, however, that the spectroscopic samples are heterogeneous,
and the face values quoted above should not be over-interpreted.
Note as well that the photometric errors of red galaxies at our magnitude limit
are typically $\sigma (V)=0.1,\  \sigma (R)=0.06,\  \sigma (i')=0.04,\  \sigma (z')=0.06$
in RXJ0153 and  $\sigma (B)=0.09,\  \sigma (V)=0.05,\  \sigma (R)=0.02,\  \sigma (i')=0.02,\  \sigma (z')=0.04$
in CL0016 (errors of blue galaxies are smaller than these values).
Thus, photo-z should work with a reasonably good accuracy ($|\Delta z|<0.1$) even at our magnitude limits \citep{kodama99}.

Using cluster galaxies only ($0.78<z_{spec}<0.88$ for RXJ0153 and
$0.50<z_{spec}<0.60$ for CL0016), we examine the accuracy of photometric redshifts
as a function of galaxy colour.
For RXJ0153, we cannot identify any bias in the photometric redshifts,
although we lack galaxies having intermediate colours ($R-z'\sim1.4$).
We find, however, that we tend to underestimate redshifts for galaxies
with intermediate colours ($V-i'\sim1.5$) by $\Delta z\sim-0.1$ in CL0016.
A possible reason for this colour dependence is that
slightly bluer galaxies at cluster redshift than those on the CMR
tend to be regarded as red galaxies at slightly lower redshifts
because of the colour--redshift degeneracy
\citep{kodama99}.
But, we need more spectroscopic data to fully address this issue.
Our photometry covers a similar rest-frame wavelength regime
for both RXJ0153 and CL0016, and we expect that we also tend to
underestimate redshifts of galaxies with intermediate colours in RXJ0153.
In both samples, photometric redshifts of red galaxies are found to be fairly accurate,
though a small offset $z_{photo}-z_{spec}=+0.02$ is seen.

Galaxies outside of a certain photometric redshift range around the cluster redshift
are regarded as fore-/background galaxies.
The choice of the redshift range is a trade off between
the completeness of cluster galaxies and the contamination of fore-/background galaxies.
If we adopt a small redshift range, the contamination will be reduced,
but a price to be paid is a selection bias towards red galaxies.
If a wide redshift range is adopted, the selection bias will be reduced
at the cost of increasing an amount of the contamination.
Since we are interested in galaxy properties, especially colours,
we prefer to construct an unbiased sample.
For this purpose, we adopt $z_{cl}-0.12<z_{\rm phot}<z_{cl}+0.05$,
where $z_{cl}$ is a cluster redshift.
This selection criteria will eliminate the colour selection bias well,
while maintaining an amount of the contamination to be minimal
(Figure \ref{fig:photoz_vs_specz}).
To be specific, we adopt $0.42<z_{\rm phot}<0.60$ for CL0016
and $0.71<z_{\rm phot}<0.88$ for RXJ0153.
Such an asymmetric redshift ranges are taken because of the possible asymmetric
photo-z error distributions (see Fig. \ref{fig:photoz_vs_specz}).

It should be noted that these redshift ranges are different from those adopted
in \citet{kodama05}.
They adopted narrower redshift ranges to enhance large-scale structures
by tracing, primarily, red galaxies.

\subsection{Statistical Contamination Subtraction}
Even though the photometric redshift is effective to reduce the fore-/background
contamination, contamination still remains at a non-negligible level
within our photometric redshift ranges.
We construct a control field sample and statistically subtract such remaining contamination
on the basis of the galaxy distribution on CMDs.
Note that we do not apply any contamination subtraction in the SDSS sample since
galaxies are spectroscopically observed.

Here we describe only the essence of the procedure of the statistical subtraction.
See Appendix \ref{app:field_sub} for details.
We adopt a modified procedure of \citet{kodama01b} and \citet{pimbblet02}.
In brief, the distribution of control field galaxies on the CMD is used
as a probability map of the contamination.
The control field sample is defined as low-density regions
in each field of RXJ0153 and CL0016, as shown later.
The choice of the control field is arbitrary, but our results do not strongly
rely on a particular choice of the control field.
We select two separated regions in each field to minimize cosmic variance.
The average galaxy density in the selected regions is denoted as $\Sigma_{control}$.
The subtraction of the remaining contamination is done on the CMD
using a Monte-Carlo method.
We statistically subtract the galaxy distribution on the CMD of
the control field from that of the target field.

\subsection{Concerns about the Contamination Subtraction}
Although we carefully subtract the contamination, uncertainties
arising from the subtraction cannot be ignored.
In this subsection, we briefly address the robustness of our conclusions presented below.

In the following sections, we discuss the fraction of red galaxies.
As shown above, photometric redshifts of red galaxies are more accurate
than those of blue galaxies, and we may tend to miss blue galaxies.
Therefore, the real fraction of red galaxies would be smaller than we observe.
This will strengthen our conclusions, the observed density dependence of galaxy
colours and evolutionary trends.
We cannot, however, evaluate the amount of missing blue galaxies
with the data at hand.

Errors in the statistical subtraction is also a concern (e.g., over/under subtraction).
However, taking advantage of the wide field of view, we can take wide areas 
for our control field sample, namely $\sim270\rm\ arcmin^2$ and
$\sim210\rm\ arcmin^2$ in RXJ0153 and CL0016, respectively.
Therefore cosmic variance must be averaged to some extent
and is not expected to be a serious problem.
We repeated Monte-Carlo runs of statistical subtraction many times 
and confirmed that trends we discuss later are seen in most realization.
In other words, we discuss secure results only.
Although the robustness of our contamination subtraction
must be confirmed later spectroscopically, it is unlikely that our results
significantly suffer from the uncertainties in the contamination subtraction.

%
%
\section{Definitions of Environmental Parameters and Derivation
of Rest-Frame Quantities and Stellar Mass}
\label{sec:parameters}

Before presenting our results, we summarize various photometric and environmental
parameters used in this paper.
We begin by introducing environmental parameters used to characterize local and global environments.
We then describe galaxy properties such as magnitude, colour, and stellar mass.

\subsection{Environmental Parameters}
\subsubsection{Local Density}
In this paper, we use nearest-neighbor density to characterize environment mainly because
it is a frequently used indicator and comparisons with other studies can be made directly.
In RXJ0153 and CL0016, all the galaxies in the selected photometric redshift range are
projected onto the redshift of the main cluster, and density is estimated from the distance
to the 10-th nearest galaxy from the galaxy of interest and is denoted as $\Sigma_{10\rm th}$.
We use a circular aperture in the density calculation.
This density is called local density hereafter.
It should be noted that local density is actually a surface density.
Galaxies that reach the edge of our field of view before finding their 10-th nearest galaxies
are not used in the analysis since the local density of such galaxies is not correctly estimated.
Local density for the two high-$z$ clusters is calculated in both physical and comoving scales.

In the SDSS, local density is estimated in a similar manner as \citet{balogh04b}.
In brief, galaxies within $\pm1000\kms$ in the line-of-sight velocity space
from the galaxy of interest are projected onto the redshift of the central galaxy,
and local density is defined from the distance to 5-th nearest neighbor.
When counting galaxies, we use only those brighter than $M_V=-19.5$,
which is volume-limited out to $z=0.065$ (our maximum redshift range).
This density is denoted as $\Sigma_{5\rm th}$.
Although we use the distance to the 5-th nearest neighbor, our results are essentially unchanged
if we use the 10-th nearest neighbor as used for RXJ0153 and CL0016.
Since the contamination of fore-/background galaxies is very small in the SDSS sample,
we prefer to adopt a smaller number so that density represents more 'local' environments.
The fiber-collision problem may affect our density estimates in high-density environments.
However, the effect is not significant \citep{tanaka04}.
Local density for the SDSS is calculated in the physical scale only.
There is, however, little difference between the physical and comoving density since
galaxies lie at very low redshifts.

\subsubsection{Global Density}
\label{sec:global_density}
Global density is defined as the surface galaxy density in a fixed radius of 2 Mpc
around the central galaxy.
Galaxies are projected onto the cluster redshift (RXJ0153 and CL0016)
or onto the redshift of the galaxy (SDSS) in question, in just the same manner as local density.
Since we aim to characterize global environment, global density is evaluated as
a comoving density.
By combining local density with global density, we can separate poor groups from
rich clusters quantitatively.
The effectiveness of this method is described in Appendix \ref{app:global_density}.
In what follows, global density is denoted as $\Sigma_{global}$.

\subsection{Magnitudes, Colours, and Stellar Masses}

\subsubsection{Rest-frame Magnitudes and Colours}

For RXJ0153 and CL0016, we derive rest-frame absolute $V$-band magnitude
and $U-V$ colour from the observed magnitudes and colours using the model templates
of the photometric redshift code \citep{kodama99}.
The conversion applied in the observed quantities to derive
the rest-frame quantities is small.
Within the selected redshift ranges, the effect of galaxy evolution is
estimated to be $\Delta M_V<0.2$ and $\Delta (U-V)<0.05$ for both RXJ0153 and CL0016.
As for the SDSS sample, we estimate the rest-frame $V$ and $U-V$
using the code by Blanton et al. (2003c, v3\_2).
We recall that, for conventional reasons, we use the Vega-referred system
in the rest-frame $V$ magnitude and the $U-V$ colour.
Table \ref{tab:rest_summary} summarizes the limiting magnitudes for the three samples.

\subsubsection{Stellar Masses}

We derive approximate stellar masses of galaxies.
For each sample, the stellar mass to light ratio ($M_*/L$) and $M_*$ are
derived from the model templates used in the photometric redshift code.
Our $L$ is defined at the rest-frame $\sim V$-band, and so the estimates of
$M_*$ are affected by on-going/near-past star formation activities.
It is expected that our stellar mass estimates are relatively accurate
for red galaxies, since red galaxies are not actively forming stars.
However, stellar masses of blue galaxies cannot be reliably determined.
We find that $M_*/L$ ratios span a factor of $\sim4$ depending on the colour
of galaxies, and we can get only rough estimates in stellar mass
although we correct for the mass-to-light ratio based on the colours
(SED fitting).
Moreover, stellar mass is a model-dependent quantity, since mass-to-light ratio
depends on stellar initial mass function (IMF).
Our $M_*$ estimate is based on the \citet{kodama97} population synthesis model,
and we assume the IMF of $x=1.10$ for the elliptical models and
$x=1.35$ for the disk models with the mass range of $0.1M_\odot - 60M_\odot$.
The limiting stellar masses that we can trace completely
are shown in Table \ref{tab:rest_summary}.

\begin{table}
\begin{center}
\begin{tabular}{lcc}\hline
sample  & $M_{V, lim}$ & $M_{*,lim}/M_\odot$\\
\hline
SDSS    & $-17.5$     & $4\times10^{9}$\\
CL0016  & $-18.0$     & $5\times10^{9}$\\
RXJ0153 & $-18.0$     & $4\times10^{9}$\\
\hline
\end{tabular}
\caption{
A list of the $V$ limiting magnitudes (Vega) and the limiting stellar masses of
the three samples.
}
\label{tab:rest_summary}
\end{center}
\end{table}

%
%
\section{Colour-Density Relations}
\label{sec:colour-density}

We apply photometric redshift technique and discover large-scale structures
around both RXJ0153 and CL0016 clusters.
Details are described in \citet{kodama05}.
Due to the wide-field coverage of the Suprime-Cam, we obtain a wide variety of
environments, i.e., sparse fields, poor groups, and rich clusters.
Here we examine the relationship between galaxy colours (star formation rates) and environments.
We stick to local environments in this section.
Effects of global environments are examined later.
First, we characterize environment by surface galaxy density (local density).
Next, environment is defined on the basis of surface mass density.

\subsection{Dependence on Surface Galaxy Density}

Based on data from large surveys of the local Universe, \cite{lewis02} and \cite{gomez03}
showed that galaxy star formation begins to decline sharply, in a
statistical sense, at a certain local density.
In what follows, this sharp decline is referred to as 'break',
and the density where 'break' is seen is referred to as break density.
Non-star-forming galaxies dominate regions above the break density, whereas
star-forming galaxies are the dominant population below the break density.
Based on the data from the SDSS, \cite{tanaka04} showed that the break is seen only for
galaxies fainter than $M_r^*+1$, and brighter galaxies show no clear break in the local Universe.
Following this work, we examine the environmental dependence of star formation
for bright and faint galaxies separately.
Galaxies brighter than $M_V^*+1$ are defined as bright galaxies, and those
fainter than that limit are defined as faint galaxies
(the value of $M_V^*$ at each redshift is given in \S\ref{sec:luminosity_function}).

\begin{figure*}
\begin{center}
\leavevmode
\epsfxsize 0.45\hsize \epsfbox{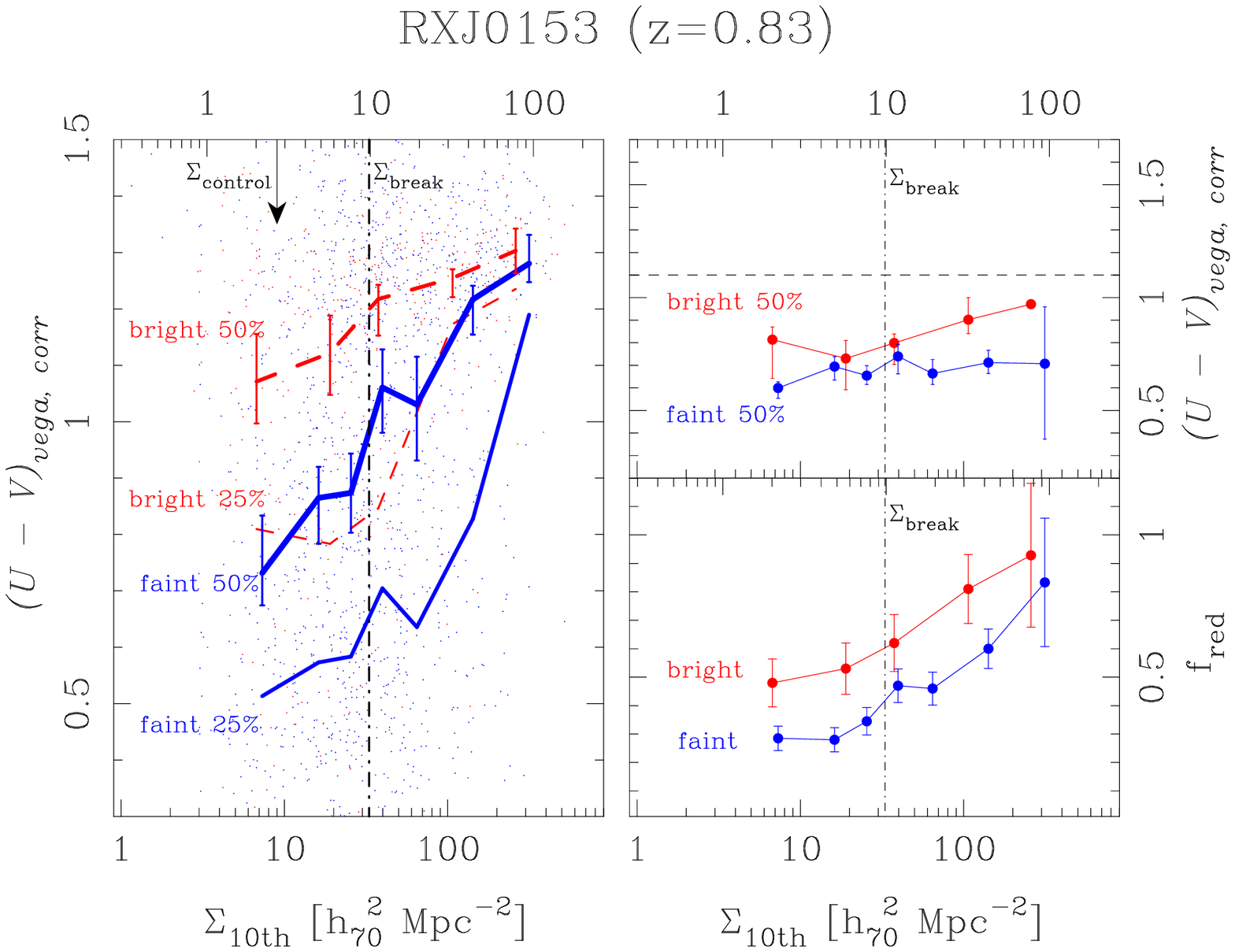}\hspace{0.5cm}
\epsfxsize 0.45\hsize \epsfbox{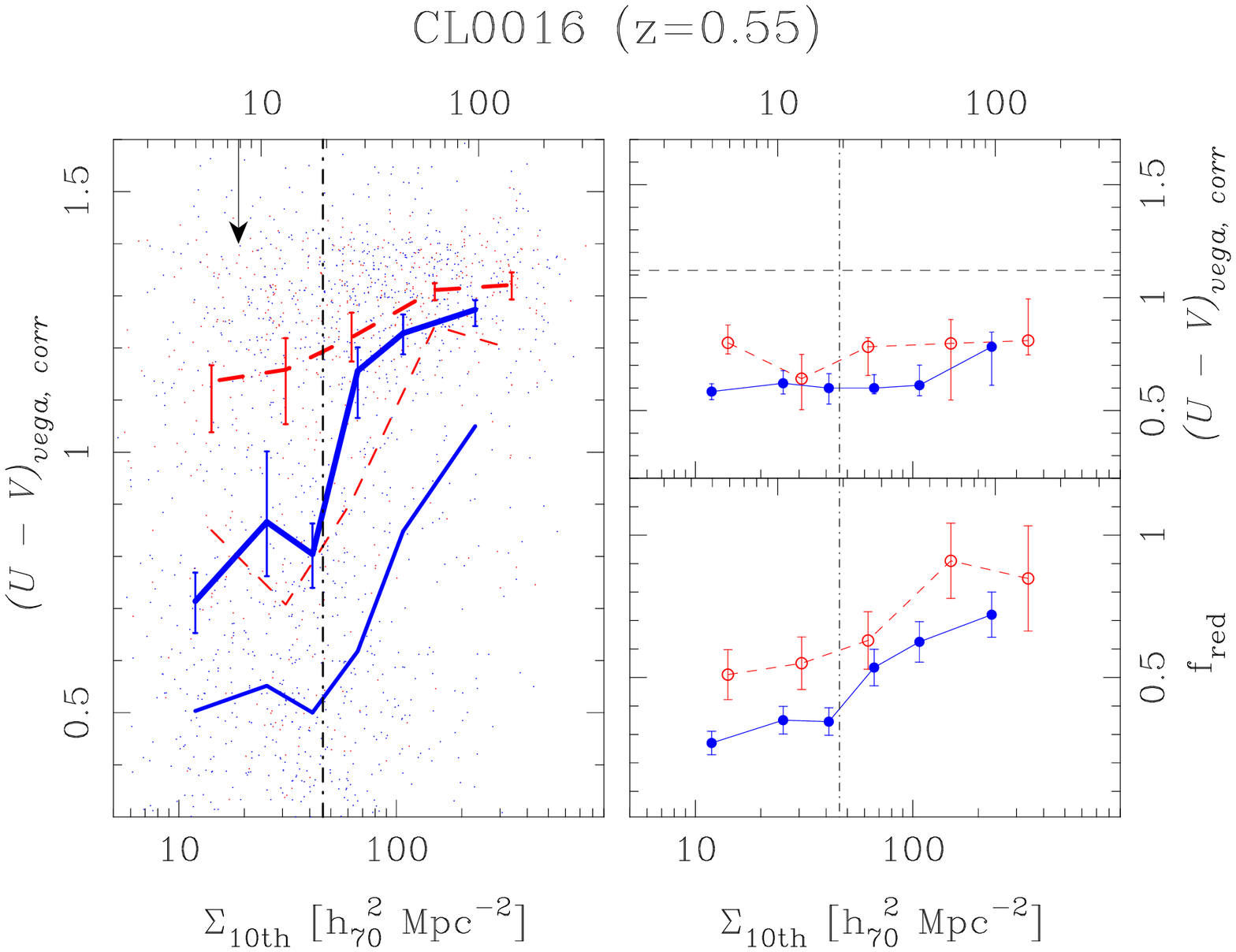}\\
\epsfxsize 0.45\hsize \epsfbox{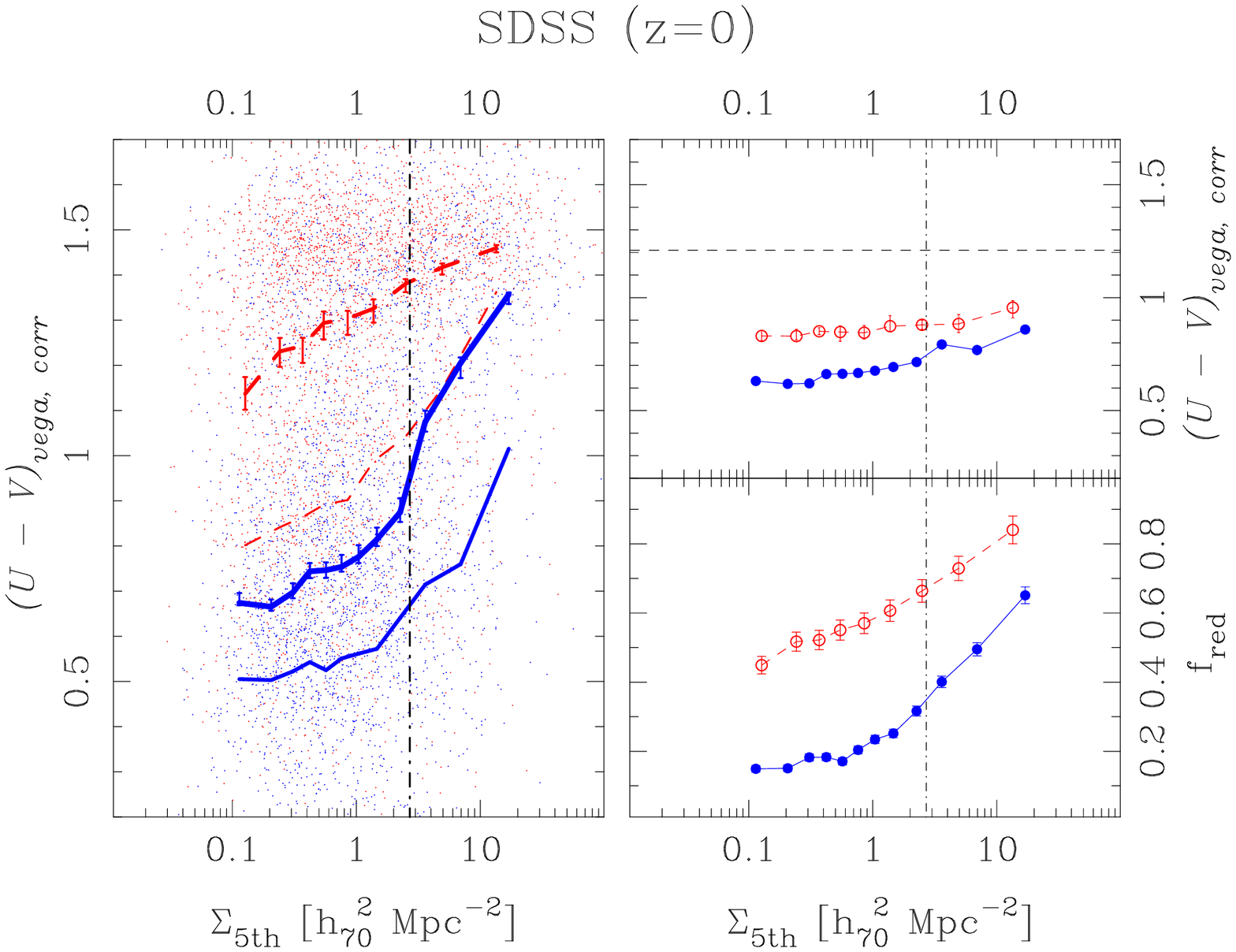}
\end{center}
\caption{
The colour-density relation in RXJ0153 ({\it top-left}),
CL0016 ({\it top-right}), and SDSS ({\it bottom}).
These plots show one realization of our Monte-Carlo run for
the statistical contamination subtraction.
Note that no contamination subtraction is performed in the SDSS plot.
{\it Left panel in each plot}: 
The $(U-V)_{corr}$ colour plotted against local density.
The lines show the median and 25-th percentile of the distribution of bright (dashed line)
and faint (solid line) galaxies as noted in the RXJ0153 panel.
The median line is associated with bootstrap 90\% intervals
as shown by the error bars.
Densities are expressed in the comoving density (top ticks) and
in the physical density (bottom ticks) and are contamination corrected
(i.e., shifted leftward by $\Sigma_{control}$).
The vertical lines show the break density of each cluster.
The arrow indicates $\Sigma_{control}$, where a half of galaxies are statistically subtracted.
In the SDSS plot, local density is shown in a physical scale only.
For clarity, one tenth of all the SDSS galaxies are randomly selected and plotted.
Each bin contains 100 bright / 200 faint galaxies 
in RXJ0153 and CL0016, and
1000 bright / 2000 faint galaxies in the SDSS plot.
{\it Top-right panel in each plot}: 
The $(U-V)_{corr}$ colour plotted against local density for
blue [$U-V<(U-V)_{CMR}-0.15$] galaxies.
The lines show the median of the distribution of bright/faint galaxies
as noted in the RXJ0153 panels.
The associated error bars are the bootstrap 90\% intervals.
The horizontal line means $(U-V)_{CMR}-0.15$.
The vertical line shows the break density of each cluster.
{\it Bottom-right panel in each plot}:
The fraction of red galaxies plotted against local density.
The meanings of the lines are given in the RXJ0153 panels.
The errors are based on the Poisson statistics.
}
\label{fig:colour_density}
\end{figure*}

Figure \ref{fig:colour_density} shows our results.
In the figure, the $U-V$ colour is corrected for the slope of the CMR so that
galaxies on the CMR have the same colour as that of a $M_V^*$ galaxy and
is denoted as $(U-V)_{corr}$ (i.e., CMR is transformed into a horizontal sequence).
The faint galaxies (solid line) show a break, prominent change in their $(U-V)_{corr}$ colours,
at the densities marked by the dot-dashed lines in the figures.
It is interesting that the break in galaxy colours is seen at all redshifts examined here.
In contrast, the bright galaxies (dashed line) do not show such a strong break,
especially the median lines.
The $(U-V)_{corr}$ colours of bright galaxies are systematically redder than
those of faint galaxies at any local density.

\begin{figure}
\begin{center}
\leavevmode
\epsfxsize 1.0\hsize \epsfbox{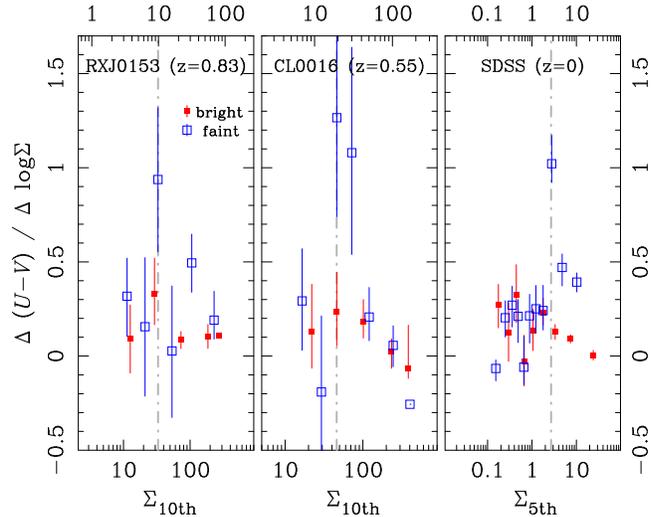}
\end{center}
\caption{
The slope of the colour-density relation [ $\Delta (U-V) / \Delta\log\Sigma$ ]
as a function of local density along the median loci in Fig.~\ref{fig:colour_density}.
The top/bottom ticks are comoving/physical densities as in Fig.~\ref{fig:colour_density}.
The panels show RXJ0153, CL0016, and SDSS (from left to right).
The filled/open symbols show bright/faint galaxies, respectively.
The vertical dot-dashed line means the break density.
The errors are estimated from the bootstrap resampling.
}
\label{fig:break_density}
\end{figure}

In order to quantify the break feature, we measure 
the slope of the colour-density relation
using the median lines in Fig. \ref{fig:colour_density}
as a function of local density for bright and faint galaxies separately.
The result is shown in Fig.~\ref{fig:break_density}.
The faint galaxies show a strong change in slope at the densities
shown by the dot-dashed lines, while bright galaxies show a much weaker
change there.
Here we define the break density at which we see the strong colour change
in faint galaxies, as marked
in Figs. \ref{fig:colour_density} and \ref{fig:break_density}.

We change the threshold magnitude used for separating bright/faint galaxies
and check  how the presence of the break density of each cluster
depends on the magnitude of galaxies.
It is found that galaxies brighter than $M_V^*+1$ do not show a prominent break,
while those fainter than that limit show a break at the same density.

There are two possible effects that cause the break.
One is that the fraction of red galaxies relative to blue galaxies
begins to increase above the break density.
The other is that blue galaxies become systematically redder (but bluer than galaxies on the CMR)
above the break density.
These possibilities are investigated in the right panels of each plot in Figure \ref{fig:colour_density}.
Red and blue galaxies are separated at $(U-V)_{CMR}-0.15$.
In RXJ0153 and CL0016, the $(U-V)_{corr}$
of blue galaxies does not change with density,
while the fraction of red galaxies strongly changes with density.
This means that the break is caused by the change in the population fraction of red galaxies.
As for the SDSS plot, the $(U-V)_{corr}$ of faint blue galaxies becomes
systematically redder above the break density.
On the other hand, the $(U-V)_{corr}$ of bright blue galaxies does not strongly change with density.
The fraction of red galaxies is found to strongly depend on density.
Therefore, we conclude that, in the SDSS plot, the break is caused by both effects.

The break densities are used to define environments in \S\ref{sec:env_def}.
We note that a small change in the break density has no significant effect on our results.
Note as well that the break density at $z=0.55$ and $z=083$ is 3-5 times higher
than the control field density, and thus the uncertainty in the statistical
contamination subtraction is not a concern.
As described in \S 3.3, photometric redshifts may miss a fraction of blue galaxies.
This will strengthen the observed break since the break is primarily driven by the change
in the population fraction of red galaxies relative to blue galaxies.
Further discussion on the break density of each cluster is made in \S 9.2.

\subsection{Dependence on Surface Mass Density}

\citet{gray04} first presented the environmental variation of galaxy colours
as a function of surface mass density.
Inspired by this work, we discuss colours of galaxies as a function of
surface mass density in this subsection.
This investigation is particularly interesting since surface galaxy density,
on which our analysis in the previous subsection is based, represents
density of luminous matter around the cluster, while the surface mass density
estimated via the weak-lensing analysis represents total mass including dark matter.
These two densities do not necessarily agree with each other,
and comparisons between the dependence on galaxy density and that on mass density
will give us a hint of physical mechanisms that affect galaxy  properties.

The weak lensing mass reconstruction of RXJ0153 is described in detail in
\citet{umetsu04}.
The lensing convergence $\kappa$ is related to the surface mass density
$\Sigma_{\kappa}$ by
$\kappa(\vec{\theta}) =
\Sigma_{\kappa}(\vec{\theta}) / \Sigma_{\kappa,{\rm crit}}^{\rm eff}$.
As a fiducial value, \citet{umetsu04} adopted 
$\Sigma_{\kappa,{\rm crit}}^{\rm eff}=3.1\times  10^{15} 
h_{70} M_{\odot} {\rm Mpc}^{-2}$.

\begin{figure}
\begin{center}
\leavevmode
\epsfxsize 1.0\hsize \epsfbox{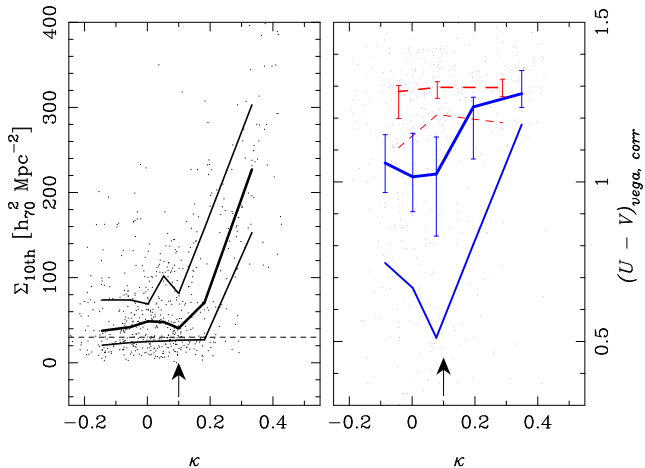}
\end{center}
\caption{
{\it Left:}
Relationship between the surface galaxy density (local density) and
normalized surface mass density ($\kappa$) in RXJ0153.
The lines show the median and quartiles (25\% and 75\%) of the distribution.
The horizontal dashed line means the break surface galaxy density.
The arrow indicates $\kappa=0.1$ where $S/N$ of the mass map is unity.
{\it Right:} The $(U-V)_{corr}$ colour plotted against $\kappa$.
The meanings of the lines are the same as Fig \ref{fig:colour_density}.
The arrow indicates mass density of $S/N=1$.
}
\label{fig:density_vs_mass}
\end{figure}

We plot in the left panel of Figure \ref{fig:density_vs_mass}
the surface galaxy density against 
the lensing convergence $\kappa(\vec{\theta})$. 
A positive correlation is found between $\kappa$ and
galaxy surface density, especially at high densities
with $\kappa \gsim 0.1$.
On the other hand, no clear correlation can be seen in the low
density regime. 

In the right panel, the $(U-V)_{corr}$ colour is plotted against $\kappa$.
Although it is not as clear as seen in Figure \ref{fig:colour_density},
there is a hint of a break in the $(U-V)_{corr}$ colour at $\kappa\sim0.1$.
This threshold corresponds to
 $\Sigma_{\kappa} \sim 3 \times10^{14}\ 
 {\rm M_\odot\ }h_{70}{\rm Mpc^{-2}}$ in physical units. 
\citet{gray04} found a similar break at the surface mass density of
$\Sigma_{\kappa}\sim 3.6\times10^{14}\ {\rm M_\odot\ }h_{70}{\rm Mpc^{-2}}$,
which is consistent with our estimate.
However, this apparent threshold density of $\kappa\sim 0.1$ is comparable
to the {\it rms} noise level in the reconstructed $\kappa$ map,
$\sigma_{\kappa}\simeq 0.10$. Therefore, the underlying mass density threshold
can be smaller than what we obtained, $\kappa\sim 0.1$.
It is therefore premature to say if galaxy properties are
more strongly related to galaxy density than to mass density.
We note that \citet{jee04} recently reported that distribution of galaxies, mass, and
intracluster medium are all different in this cluster suggesting on-going cluster merger.

Although we cannot draw a firm conclusion on this analysis,
this is potentially an interesting way of investigating environmental
dependence of galaxy properties.
If, for example, galaxy-galaxy interactions are the main driver of the environmental dependence,
we expect to see stronger dependence on galaxy density than on mass density.
On the other hand, if interactions with cluster potential is the main driver,
we expect a stronger dependence on mass density.

%
%
\section{Definitions of Field, Group, and Cluster Environments}
\label{sec:env_def}

\begin{figure*}
\begin{center}
\leavevmode
\epsfxsize 0.7\hsize \epsfbox{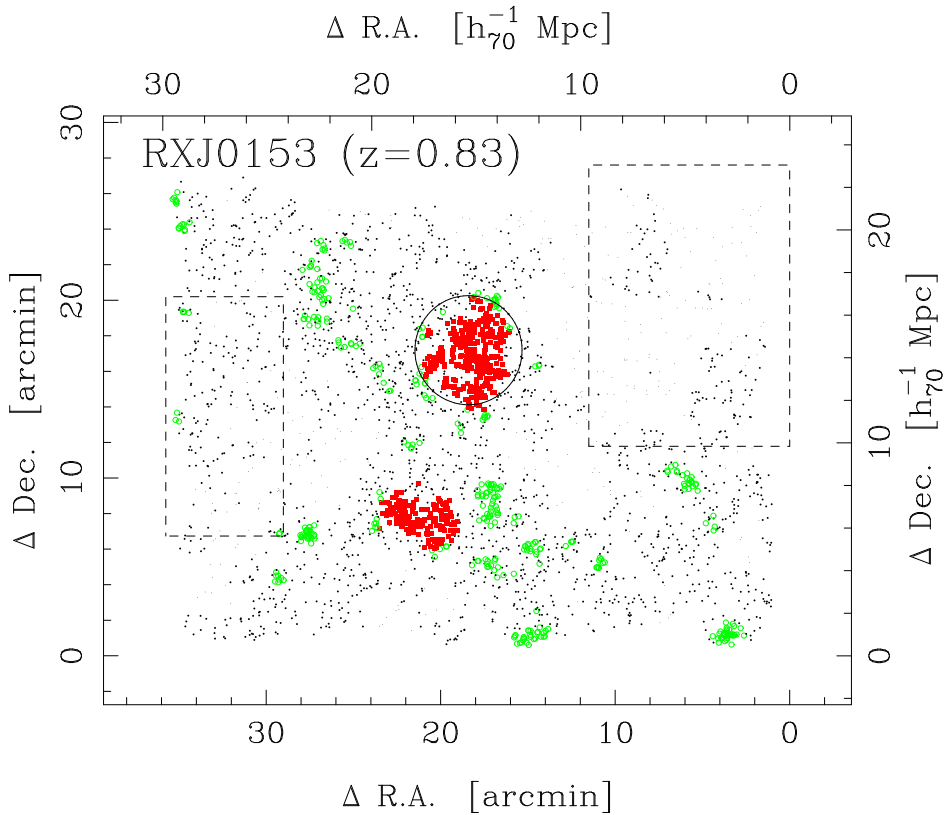}
\end{center}
\caption{
Galaxy distribution in RXJ0153. North is up.
The filled rectangles and open circles show cluster and group galaxies.
The large dots are field galaxies.
Galaxies in the lower density regions than $\Sigma_{control}$ are shown by the small dots.
Galaxies in the large dashed rectangles are used as control field galaxies,
and they are used in the statistical contamination subtraction procedure.
Galaxies too close to the edge of our field of view are not plotted
for they have no local density estimates.
The top and right ticks show the comoving scales in unit of Mpc.
The circle shows $R_{vir}$ \citep{maughan03}.
No statistical contamination subtraction is applied in this plot.
}
\label{fig:rxj0152_distrib}
\end{figure*}

\begin{figure*}
\begin{center}
\leavevmode
\epsfxsize 0.7\hsize \epsfbox{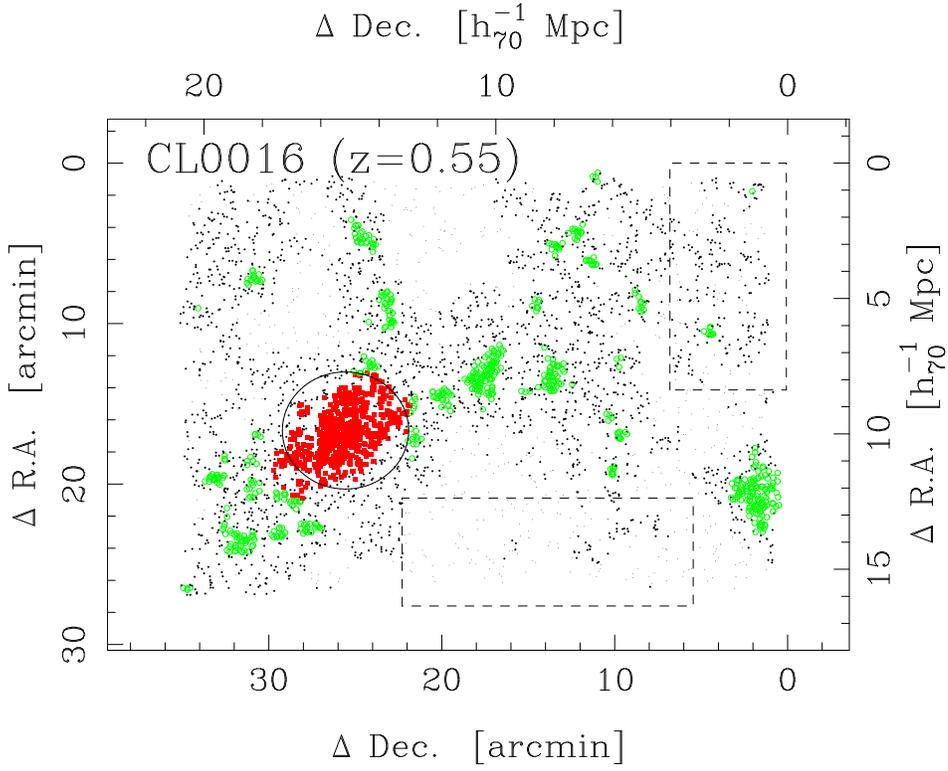}
\end{center}
\caption{
Same as Figure \ref{fig:rxj0152_distrib}, but for CL0016.
North is to the left.
}
\label{fig:cl0016_distrib}
\end{figure*}

\begin{figure*}
\begin{center}
\leavevmode
\epsfxsize 0.7\hsize \epsfbox{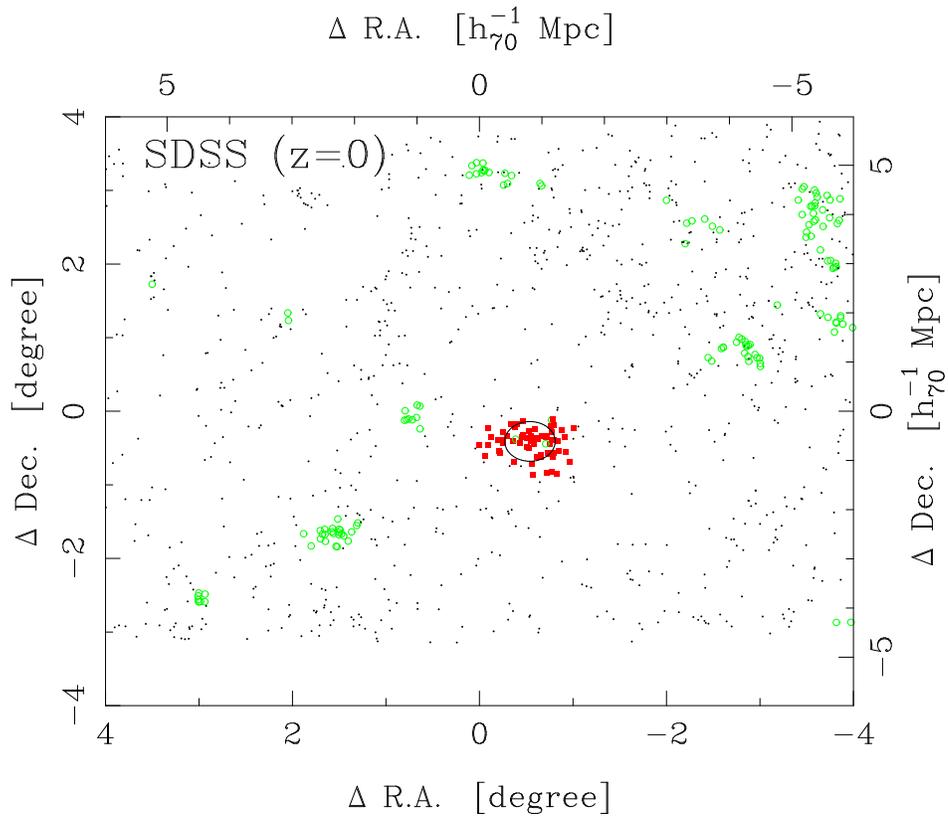}
\end{center}
\caption{
Same as Figure \ref{fig:rxj0152_distrib}, but for SDSS.
Only a patch of the sky is shown.
Galaxies are thinly populated compared with
Figures \ref{fig:rxj0152_distrib} and \ref{fig:cl0016_distrib},
but the magnitude limit here is shallower by $\sim1$ magnitude.
The dots are field galaxies.
}
\label{fig:sloan_distrib}
\end{figure*}

In order to address galaxy properties as functions of environment and time,
the definition of the environment must be quantitative and applicable at all redshifts.
For this reason, we define environments by galaxy density.
For analyses in the following sections, we define three environments:
field, groups, and clusters as shown in Figures
\ref{fig:rxj0152_distrib}, \ref{fig:cl0016_distrib}, and \ref{fig:sloan_distrib}.

We do not examine galaxies in lower density regions than $\Sigma_{control}$ 
because most of them are considered to be fore-/background galaxies.
We find in the previous section that red galaxies dominate
environments denser than the break density.
Motivated by this, we use the break density of each cluster to define environments.
Galaxies in higher density regions than $\Sigma_{control}$,
but in lower density regions than the break density ($\Sigma_{break}$)
are considered to be in the field environment
(i.e., $\Sigma_{control}<\Sigma_{local}<\Sigma_{break}$).

Galaxies in higher density regions than $\Sigma_{break}$ belong to
groups and clusters.
We now aim to investigate the effects of global environment:
are there any differences between group galaxies and cluster galaxies?
This question is particularly interesting since different 
physical mechanisms are effective in different environments.
For example, ram-pressure stripping of cold gas \citep{gunn72,abadi99,quilis00}
and harassment \citep{moore96,moore99} are expected to play a role only
in rich clusters.
Low-velocity galaxy-galaxy interactions (e.g., \citealt{mihos96}) are
expected to be effective in galaxy groups.
Strangulation (which is often referred to as suffocation, starvation or
halo gas stripping) is considered to be effective both in groups and clusters
\citep{larson80,balogh00,diaferio01,okamoto03}.

Global density defined in \S \ref{sec:global_density} is found to 
work well in separating groups from clusters.
Global density traces galaxy density over a large scale, and thus
global density is a good measure of richness of galaxy systems.
We define globally denser environment than the break density as clusters
and globally less dense environment as groups.
That is, clusters are defined by $\Sigma_{local}>\Sigma_{break}$ and
$\Sigma_{global}>\Sigma_{break}$, and groups by  $\Sigma_{local}>\Sigma_{break}$
and $\Sigma_{global}<\Sigma_{break}$.
As illustrated in Figures \ref{fig:rxj0152_distrib}, \ref{fig:cl0016_distrib}, and
\ref{fig:sloan_distrib}, our separation of environments is reasonable.
We show in each plot the virial radius of the cluster.
The virial radius of RXJ0153 is taken from \citet{maughan03},
and those of CL0016 and SDSS are evaluated from velocity dispersions
using the recipe of \citet{carlberg97}.
As seen in the plots, the break density corresponds to the outskirts
of clusters.
This is quantitatively consistent with that seen in the local Universe
where the break density corresponds to
one to two $R_{vir}$ \citep{gomez03,tanaka04}.
We note that the break density also corresponds to a density typical of isolated groups.

We are aware that, since we define environment based on galaxy colours,
we may obtain 'biased' environmental dependencies of galaxy colours.
For example, since we define cluster environment where red galaxies are abundant,
clusters are, by definition, dominated by red galaxies.
However, the fact that the break density corresponds to the outskirts of clusters
is a good justification of our definition.
Therefore, we consider that results presented below are not biased products
of the environment definition.

To sum up this section, we tabulate the definitions
of environments in Table \ref{tab:env_def}.
Based on these environments, we look into CMDs and LFs
of galaxies in the following sections.

\begin{table}
\begin{center}
\begin{tabular}{ll}\hline
Environment          & Definition\\
\hline
Field                & $\Sigma_{\rm control}<\Sigma_{\rm local}<\Sigma_{\rm break}$\\
Group                & $\Sigma_{\rm local}>\Sigma_{\rm break}$ and $\Sigma_{\rm global}<\Sigma_{\rm break}$\\
Cluster              & $\Sigma_{\rm local}>\Sigma_{\rm break}$ and $\Sigma_{\rm global}>\Sigma_{\rm break}$\\
\hline
\end{tabular}
\caption{
Definitions of environments. $\Sigma_{\rm control},\ \Sigma_{\rm local},\ \Sigma_{\rm break},$
and $\Sigma_{\rm global}$ are control field density, local density ($\Sigma_{\rm 5th}$ or
$\Sigma_{\rm 10th}$), break density, and global density.
}
\label{tab:env_def}
\end{center}
\end{table}

%
%
\section{Colour-Magnitude Diagrams}
\label{sec:colour-magnitude}

Galaxies in clusters are known to show a tight CMR (e.g., \citealt{bower92}).
This relation is observed up to $z=1$ and even beyond
\citep{kodama98,stanford98,nakata01,blakeslee04,delucia04,lidman04}.
In this section, we aim to investigate the CMDs with particular attention
to the group and field environments, which have not been
intensively studied yet especially at high redshifts.

\begin{figure*}
\begin{center}
\leavevmode
\epsfxsize 0.3\hsize \epsfbox{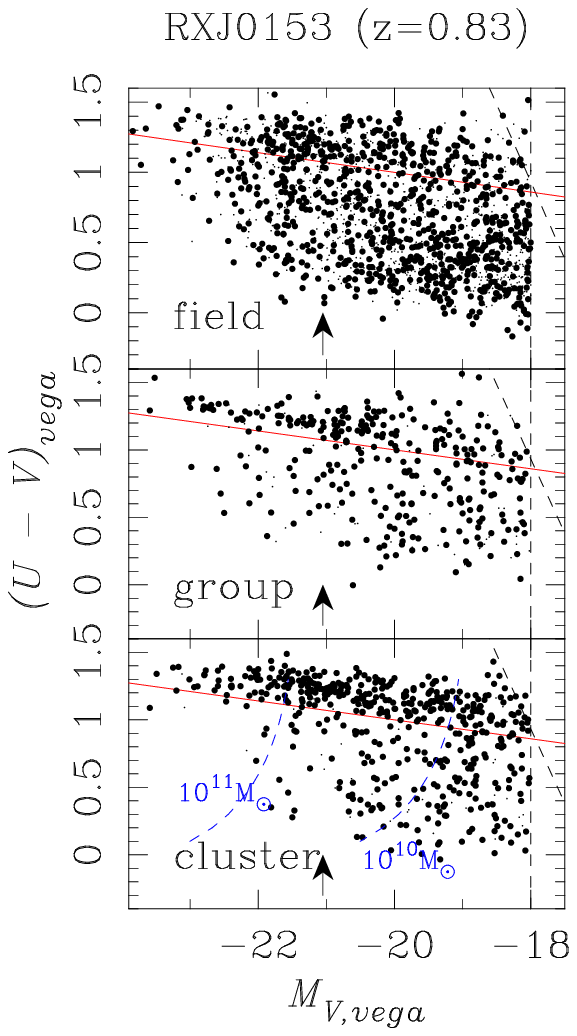}\hspace{0.5cm}
\epsfxsize 0.3\hsize \epsfbox{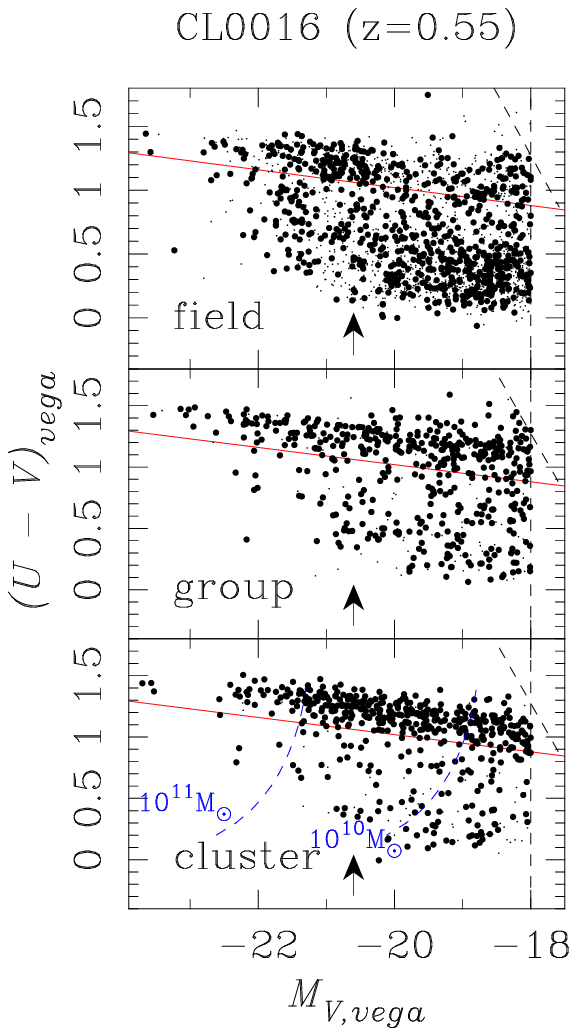}\hspace{0.5cm}
\epsfxsize 0.3\hsize \epsfbox{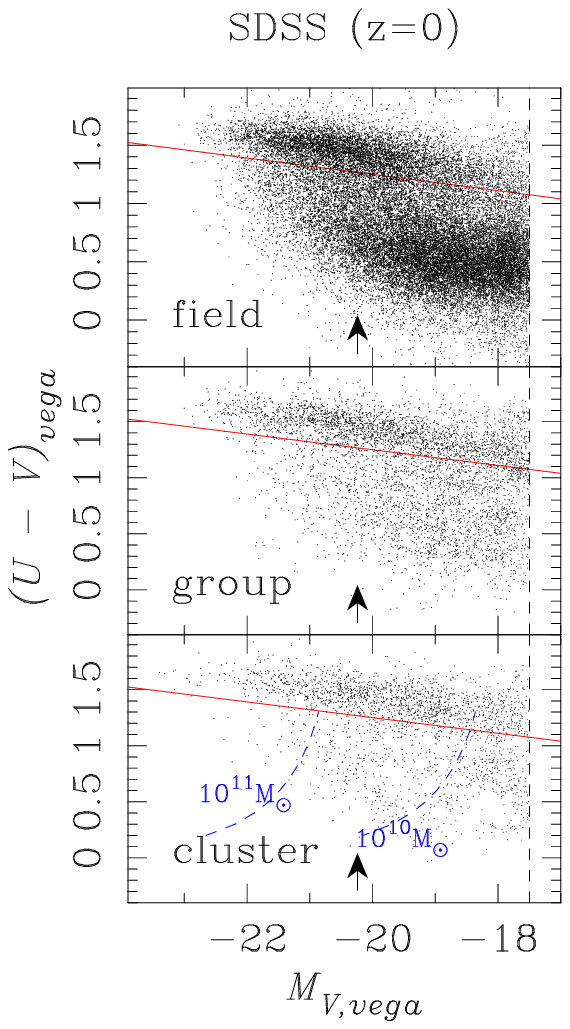}
\end{center}
\caption{
The rest-frame CMDs ($U-V$ vs. $M_V$) in RXJ0153 ({\it left}), CL0016 ({\it middle}),
and SDSS ({\it right}) are plotted.
In each plot, the panels show different environments, namely,
field, group, and cluster.
Small dots mean statistically subtracted galaxies (except the SDSS plot).
The solid line represent the CMR shifted blueward by $0.15$ magnitude.
The CMR is based on the model prediction of \citet{kodama97}.
This solid line separate red and blue galaxies used in the following analysis.
The vertical dashed line is the magnitude limit, and the slanted dashed line shows
the $5\sigma$ limiting colour.
The arrows show $M_V^*+1$, which is used to separate bright galaxies from
faint galaxies in \S \ref{sec:colour-density}.
In the cluster plot, the curved dashed lines show loci of stellar masses of
$10^{10}$ and $10^{11}\rm\ M_\odot$.
Note that, in the SDSS plot, the volume correction is applied and the corrected galaxies
are artificially plotted.
Note as well that no statistical contamination subtraction is performed in the SDSS plot.
}
\label{fig:cm_rest}
\end{figure*}

\begin{figure}
\begin{center}
\leavevmode
\epsfxsize 0.7\hsize \epsfbox{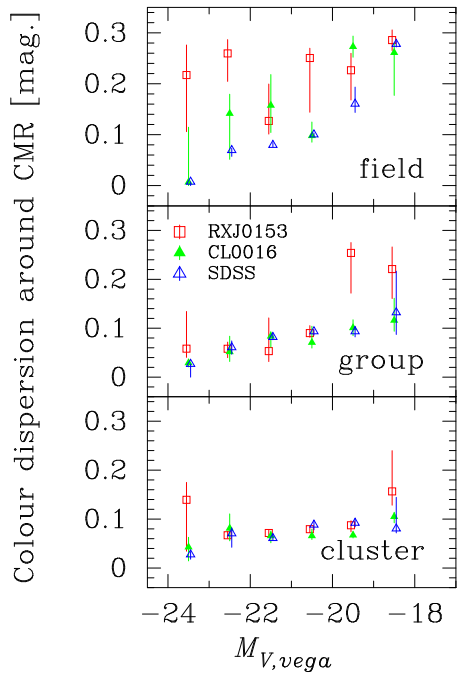}
\end{center}
\caption{
The colour dispersions around the CMR in $\Delta(U-V)$.
The meanings of the symbols are shown in the middle panel.
The error bars are estimated from the bootstrap resampling
of input catalogs.
}
\label{fig:cm_scatter}
\end{figure}

\begin{figure}
\begin{center}
\leavevmode
\epsfxsize 1.0\hsize \epsfbox{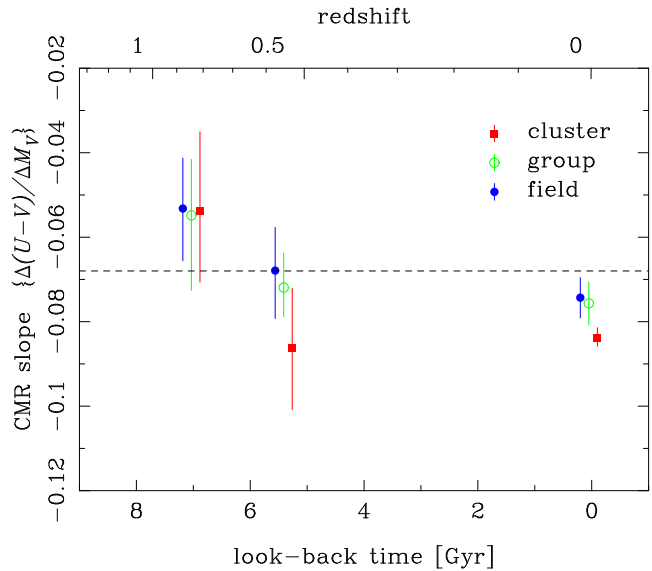}
\end{center}
\caption{
The slopes of CMRs plotted against the look-back time.
The horizontal dashed line is a prediction of \citet{kodama97} model.
The meanings of the symbols are shown in the panel.
The error bars are estimated from the bootstrap resampling.
}
\label{fig:cm_slope}
\end{figure}

Figure \ref{fig:cm_rest} shows the rest-frame CMDs.
We estimate the slope and scatter of the CMRs based on
an iterative $2\sigma$-clipping least squares fit and the results are shown
in Figures \ref{fig:cm_scatter} and  \ref{fig:cm_slope}.
Note that the CMR slopes are measured from galaxies brighter than $M_V^*+2$
which show tight relationship, while
the scatter around the CMR is measured for galaxies down
to our magnitude limits.
Errors are estimated from the bootstrap resampling of the input catalogs,
and the statistical field subtraction is performed in each run.
Since our photometric redshift cuts have some ranges ($\Delta z\sim0.18$),
a projection effect is a concern which may apparently enhance the
intrinsic colour scatter. However, the colour difference in passively
evolving galaxies within these redshift ranges are only 0.05 magnitude at
most for both CL0016 and RXJ0153 \citep{kodama98}, therefore this
effect is small compared to the amount of scatter we discuss here
($>$0.1 mag.).
We see five very interesting trends in these figures.

Firstly, we find probable onset of the build-up of the CMR in field regions.
In the field environment in RXJ0153 at $z=0.83$, we cannot identify any clear CMR,
although a clump of bright red galaxies can be seen.
In fact, the scatter around the CMR is very large (Fig. \ref{fig:cm_scatter}),
being  consistent with no CMR.
Interestingly, however, a clear relation is seen in the same environment in CL0016
at $z=0.55$ and also in SDSS at $z=0$ particularly at the bright-end.
We find that 
the scatter around the CMR at the bright-end decreases from $z=0.83$ down to $z=0$,
while the scatter at the faint-end does not decrease clearly.
That is, the bright-end of the CMR is built-up with time,
while there is no clear CMR at the faint-end even at $z=0$.

Secondly, we find on-going build-up of the CMR in groups.
Groups in RXJ0153 show a CMR, but we find that the CMR is
getting weak at $M_V>-20$.
There are a number of red galaxies at $M_V>-20$, but they do not form a tight relation.
The scatter around the CMR shown in Fig. \ref{fig:cm_scatter}
reflects this visual impression.
At lower redshifts, group galaxies show a clear relation down to the magnitude limit.
This leads us to suggest that we are witnessing the build-up of the CMR in groups.
An emerging picture is that the CMR grows from the bright-end to the faint-end,
and not the opposite way.

Thirdly, no such on-going build-up of the CMR is seen in clusters.
We cannot see  dramatic growth of the CMR in clusters.
Cluster galaxies show a clear CMR down to the magnitude limit
at all redshifts considered here
and the scatter around the CMR is already small at high redshifts.

One may suspect that the observed build-up of the CMR is an
artifact: e.g., photometric redshifts miss a fraction of red galaxies in RXJ0153,
and it mimics the CMR build-up.
But, this is not the case.
The accuracy of photometric redshifts does not depend on environment.
Accordingly, if there were a CMR in the field environments of RXJ0153,
it would have been found because we see the clear CMR in the cluster environments.
The same argument can be applied to the faint-end of the CMR in groups.
It should be noted that our group environment is a composite of several individual groups,
and hence a variation in group properties is not a major concern.
Note as well that photometric redshifts are reliable for red galaxies
(see Figure \ref{fig:photoz_vs_specz}).
Therefore, we suggest that the observed build-up is real.

Fourthly, the slope of the CMR does not change with environment.
The slope of the CMR in the field environment
is found to be similar to those in groups and clusters.
This is quantified in Figure \ref{fig:cm_slope}.
There is no convincing evidence that the CMR slope depends on environment.
Note that galaxy colours are measured within different aperture sizes at
different redshifts:
Petrosian aperture in SDSS and $2''$ aperture in CL0016 and RXJ0153.
CMR slopes change with aperture sizes \citep{bower92}, and
we do not discuss the evolution of the CMR slope here.
Although CMR seems to be built-up at different epochs in different environments,
this similarity of the CMR slope may be expected if the slope is primarily caused
by the metallicity effect rather than the age effect \citep{kodama97,stanford98}.
The observed CMR build-up suggests that the slope of CMR does not
change since the formation epoch of the CMR.

Fifthly, we confirm bimodality in galaxy colours.
Galaxy properties are known to have strong bimodal distribution in the local Universe
\citep{strateva01,blanton03a,kauffmann03,baldry04,balogh04b,kauffmann04,tanaka04}.
This bimodality is seen up to $z\sim1$ \citep{bell04}.
We confirm the bimodality although it becomes less clearly seen at higher
redshifts in all the environments.
The bimodality is particularly noticeable in the field regions of CL0016 and SDSS.
High density environments lack blue galaxies.

The highlight of this CMR analysis is that we observe the build-up of the
CMR. It seems that an evolutionary stage of the CMR build-up is different
in different environments: the cluster CMR is built first, and the CMRs of
the group and field regions are built later on.
Another interesting implication is that the bright-end of the CMR appears first,
and the faint-end is filled up later on.
We will further quantify and discuss the observed build-up in \S 9.3.

%
%
\section{Luminosity Functions}
\label{sec:luminosity_function}

Luminosity function (LF) is one of the most fundamental measures of galaxy properties.
This section studies LFs as functions of environment and time.
A LF is fitted by the Schechter function \citep{schechter76} of the form

\begin{equation}
\phi(L)dL=\phi^*\left(\frac{L}{L^*}\right)^\alpha \exp\left(-\frac{L}{L^*}\right)d\left(\frac{L}{L^*}\right),
\end{equation}

\noindent
or equivalently, per unit magnitude,

\begin{eqnarray}
\phi(M)dM=0.4\ln 10\times \phi^*10^{-0.4(M-M^*)(\alpha+1)}\nonumber\\
\exp\left[-10^{-0.4(M-M^*)}\right]dM.
\end{eqnarray}

\noindent
There are three parameters in the Schechter function:
the normalization factor $\phi^*$, the characteristic luminosity/magnitude $L^*/M^*$,
and the faint-end slope $\alpha$.
Best-fit parameters are searched via the conventional $\chi2$-minimizing statistics.
Galaxies in RXJ0153 and CL0016 are binned into 0.5 magnitude steps,
and those in the SDSS are into $0.25-0.5$ magnitude steps.
First, we examine total (=red+blue) LFs at each redshift.
Then we look into red/blue LFs in different environments and at different redshifts.

\subsection{Total Luminosity Functions}
\begin{figure}
\begin{center}
\leavevmode
\epsfxsize 1.0\hsize \epsfbox{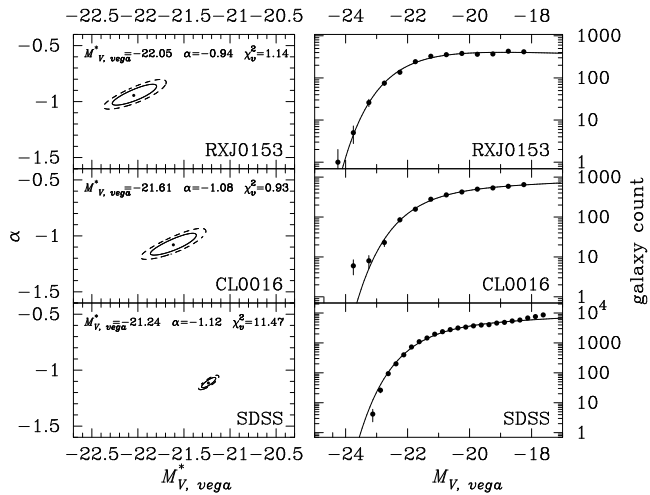}
\end{center}
\caption{
Total LFs in RXJ0153 ({\it top}), CL0016 ({\it middle}),
and SDSS ({\it bottom}).
Note that no contamination subtraction is performed here.
The left panels show the error ellipse for each LF.
The solid and dashed contours show $\chi^2_\nu=\chi^2_{\nu,best}+1$ and
$\chi^2_\nu=\chi^2_{\nu,best}+2$, where $\chi^2_{\nu,best}$ is the $\chi^2_\nu$
of the best fit.
The right panels show LFs along with the best-fit Schechter functions.
The error bars are based on the Poisson statistics only.
}
\label{fig:lf_tot}
\end{figure}

Based on the photo-$z$ selected samples of RXJ0153 and CL0016 and the spec-$z$ sample
of SDSS, the total LF of galaxies at each redshift is constructed and
shown in Figure \ref{fig:lf_tot}.
Note that no statistical contamination subtraction is performed here.

The Schechter function gives a  good fit ($\chi_\nu^2\sim1$)
to the total LF of  for RXJ0153 and CL0016, but it is not
a good fit for SDSS ($\chi_\nu^2\sim11$).
The observed SDSS LF deviates from the Schechter function at the faint-end,
and this deviation decreases the overall goodness of fit.
The deviation is possibly due to increasing contribution of dwarf galaxies.
We fit the Schechter function using galaxies with $M_V<-18.5$
and obtain $M_V^*=-21.12$ and $\alpha=-1.01$ with $\chi_\nu^2\sim4$.
We recall that the $M^*_V$ derived in Fig. \ref{fig:lf_tot} were used in
separating bright/faint galaxies in \S \ref{sec:colour-density}.
Note that a small error in $M_V^*$ has little effect on the results
obtained in that section.

Figure \ref{fig:lf_tot} clearly shows that galaxies fade with time:
$M_V^*=-22.05$ at $z=0.83$, $M_V^*=-21.61$ at $z=0.55$, and $M_V^*=-21.24$ at $z=0$.
This observed fading, $\Delta M_V^*=0.81$ mag. from  $z=0.83$ to $z=0$
and $\Delta M_V^*=0.37$ mag. from  $z=0.55$ to $z=0$, is consistent with
a passive evolution model (\citealt{kodama97}; $z_f=5$), $\Delta M_V=0.80$ mag. and
0.55 mag., respectively, within the errors.
This suggests that $M_V^*$ is primarily determined by passively evolving galaxies.
In our cosmology, $z=0.83$ and $z=0.55$ correspond to a look-back time
of 7.0 Gyr and 5.4 Gyr, respectively.

The faint-end slope $\alpha$ also seems to evolve.
The number of faint galaxies relative to bright galaxies increases with time,
$\alpha=-0.94$ at $z=0.83$ and $\alpha=-1.12$ at $z=0$.
But the local value should be taken with caution since $\alpha$ is found to
depend on the magnitude range involved in the Schechter fit.
Contribution of dwarf galaxies is likely to be significant at the faint-end.
We therefore do not try to draw any firm conclusion on the evolution of $\alpha$.

In RXJ0153 and CL0016 fields, we have rich clusters and a number of groups,
and the contribution of clusters and groups to the total LF is larger
compared with the SDSS LF.
In the SDSS sample, the field, group, and cluster galaxies comprise
80\%, 13\%, and 7\% of the total number of galaxies.
We scale the relative fraction of field, group, and cluster galaxies in RXJ0153 and
CL0016 to the SDSS values, and find that the derived Schechter parameters
show only a small change: $\Delta M_V^*=0.1$ and $\Delta \alpha=0.02$.
Thus, a different environmental mix of galaxies does not strongly change our results.

\subsection{Luminosity Functions of Red/Blue Galaxies}

\begin{figure*}
\begin{center}
\leavevmode
\epsfxsize 0.45\hsize \epsfbox{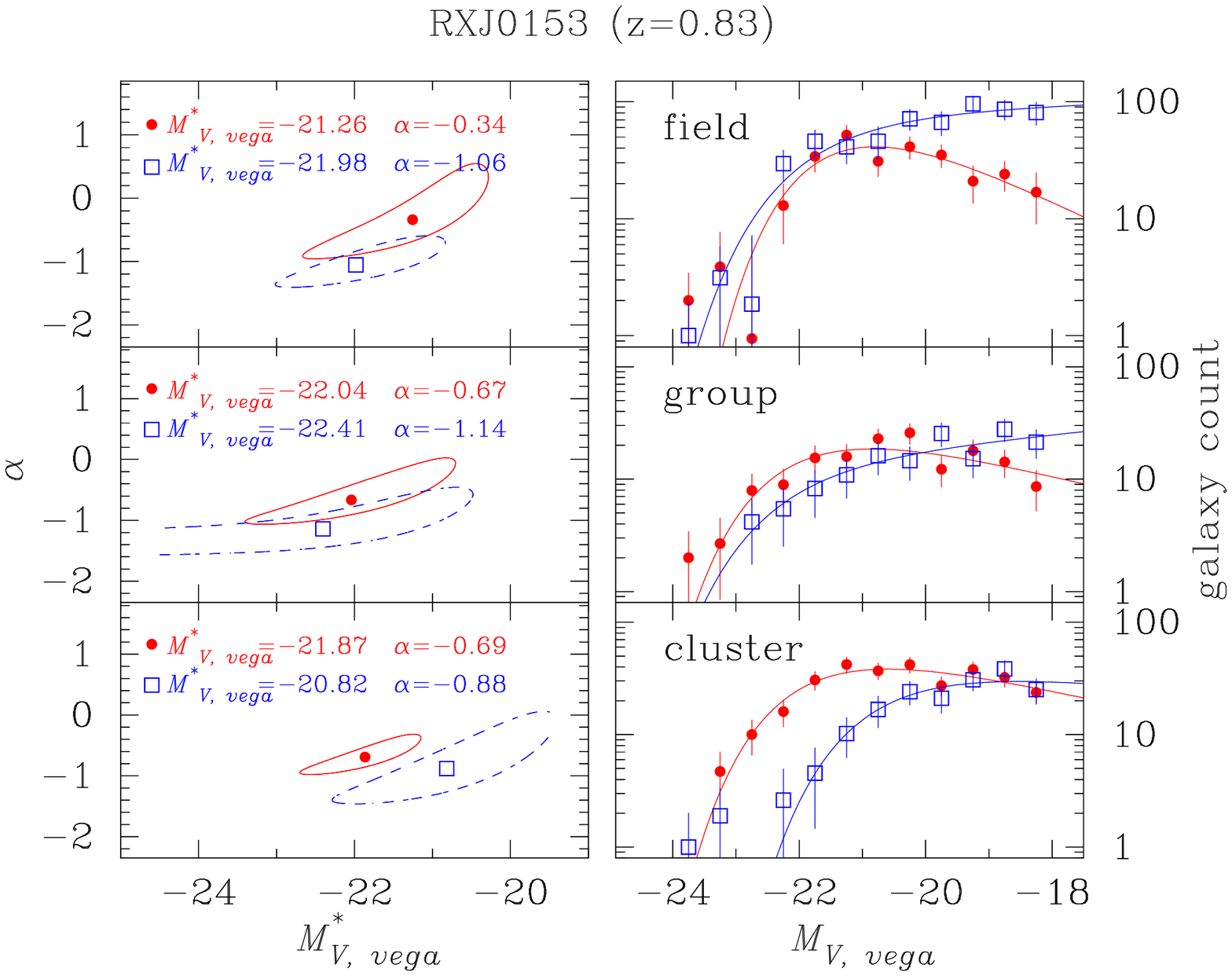}\hspace{0.5cm}
\epsfxsize 0.45\hsize \epsfbox{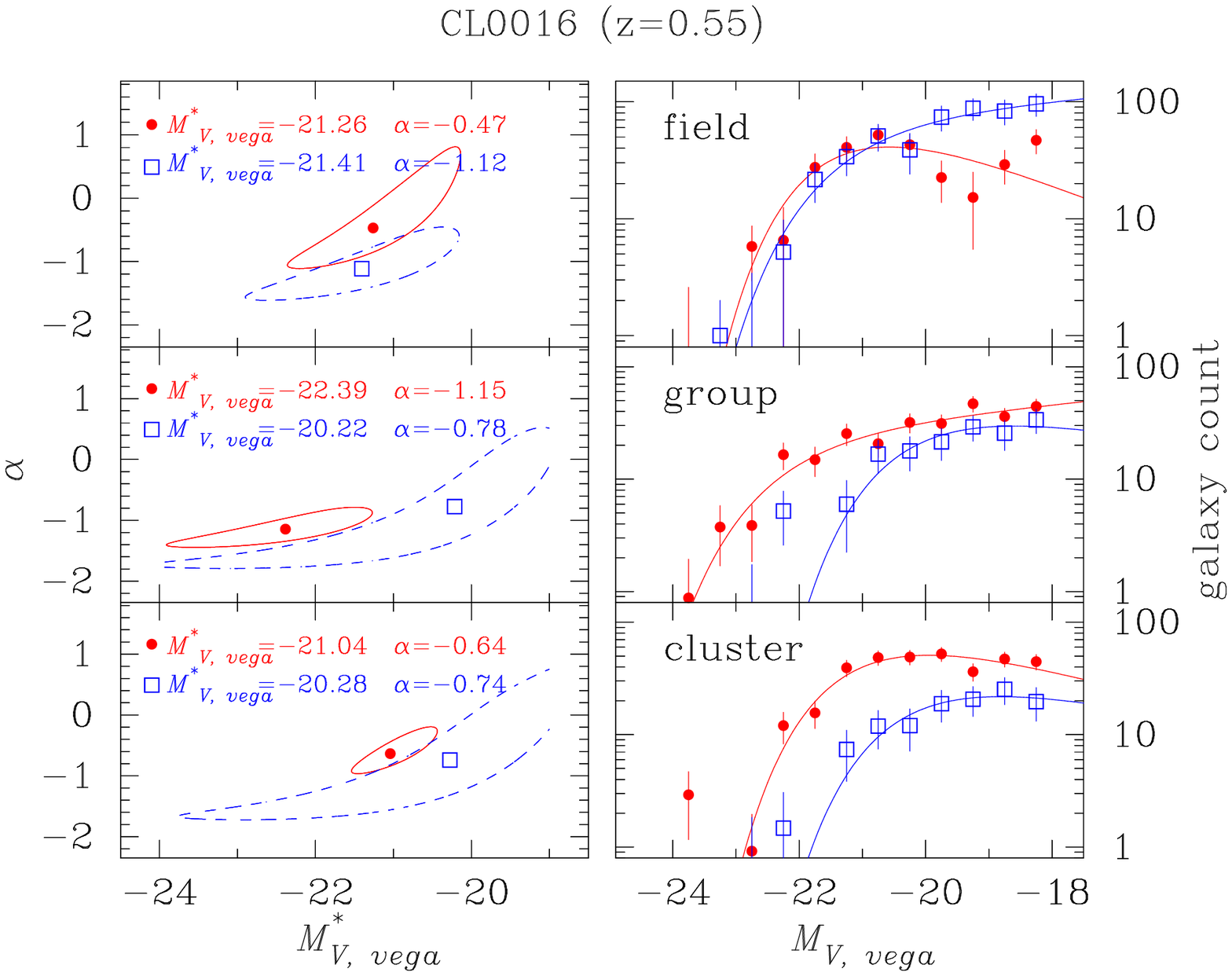}\\
\epsfxsize 0.45\hsize \epsfbox{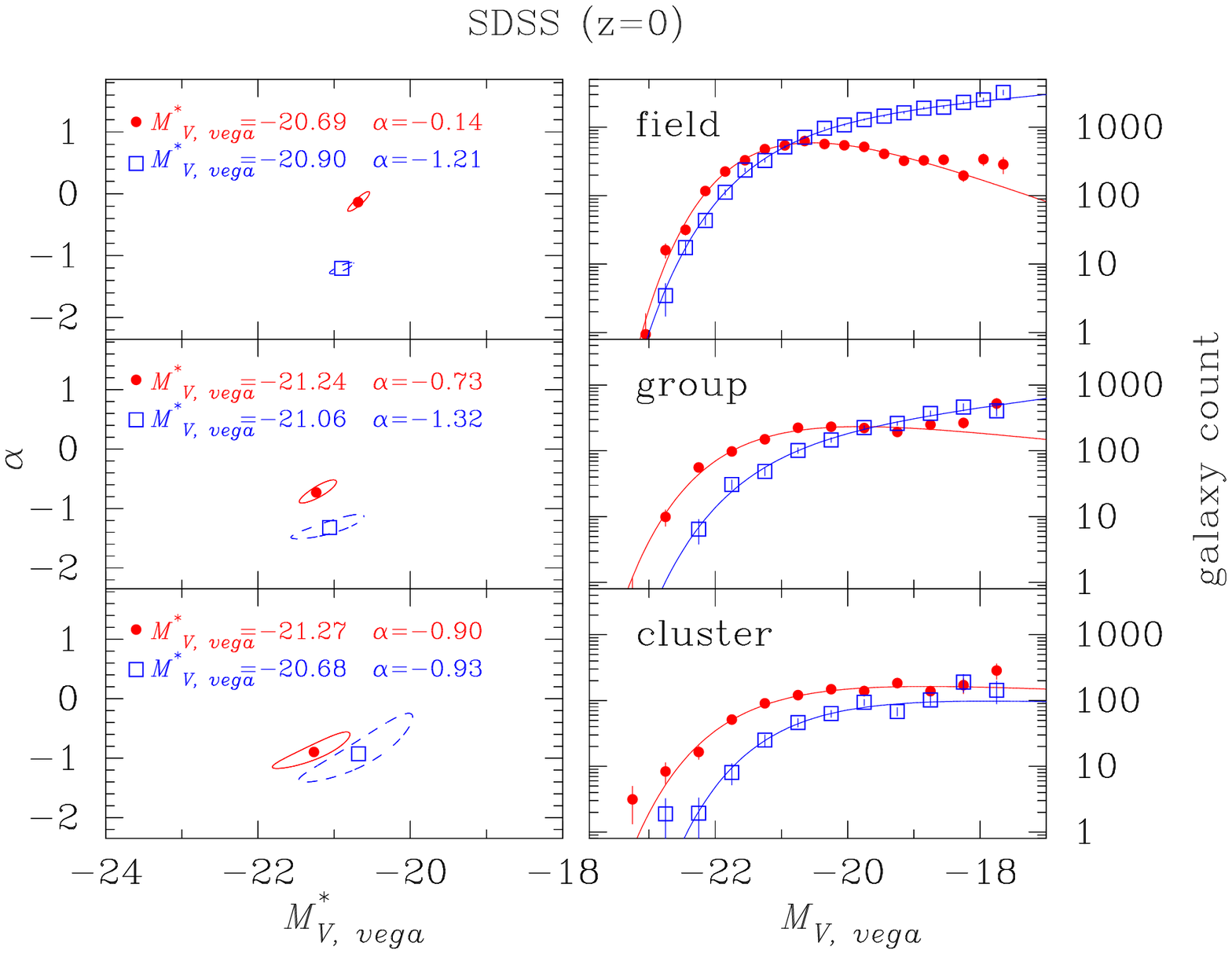}
\end{center}
\caption{
Rest-frame LFs in various environments in RXJ0153 ({\it top-left}), CL0016 ({\it top-right}),
and the SDSS ({\it bottom}).
The left panels in each plot show the error ellipses ($\chi^2_\nu=\chi^2_{\nu,best}+1$)
of the Schechter fits.
The solid and dashed lines represent red and blue galaxies defined in Figure \ref{fig:cm_rest}.
The filled points and open squares show the best-fit parameters for red and blue LFs, respectively.
The right panels show LFs.
The filled and open symbols represent red and blue galaxies.
The solid and dashed lines respectively show best-fit Schechter functions for red and blue galaxies.
The statistical contamination subtraction is carried out here.
}
\label{fig:lf_env}
\end{figure*}

The fact that galaxy properties have strong bimodality in their distribution
motivates us to investigate LFs of red and blue galaxies separately.
We define red galaxies as those having $U-V>(U-V)_{CMR}-0.15$.
Blue galaxies are defined as those bluer than this limit.
This definition is illustrated in Figure \ref{fig:cm_rest}.
It should be noted that, in the following, the statistical contamination subtraction
is performed on the LF bin -- LF bin basis.
This is different from the Monte-Carlo approach that we adopt in the previous sections.
Our results are presented in Figure \ref{fig:lf_env}.
A general trend is that red galaxies have a decreasing or flat faint-end slope
(though exceptions can be found), while blue galaxies have an increasing slope
at any redshift.
Some SDSS data points at the faint-end deviate from the best-fit Schechter function.
This is because the SDSS data at the faint-end have small weights
in the Schechter fitting due to the statistical nature of the volume correction.

Now, we focus on the LFs of red galaxies.
At $z=0$, the faint-end slope becomes gradually less steep in denser environment.
\citet{hogg03} reported, based on SDSS data, that faint red galaxies
preferentially populate in high density environments.
A similar conclusion was reached by \citet{depropris03} based on the 2dF data.
It is likely that red faint galaxies are selectively located in high density environments.
A similar trend can be seen in RXJ0153, but it is not statistically significant
given the large error ellipses.
\citet{kajisawa00}, \citet{nakata01}, \citet{kodama04}, \citet{toft04},
and \citet{delucia04} reported the deficit of red faint galaxies
in high redshift ($z>0.7$) clusters compared with local clusters.
Although we cannot confirm the trend in the Schechter parameter,
we do see the the deficit of red faint galaxies in the giant-to-dwarf ratio,
which we will discuss in the next section.
As for blue LFs, there is no convincing evidence that blue LFs strongly depend
on environment and time.

%
%
\section{Discussion}
\label{sec:discussion}

\subsection{Possible Peculiarity of CL0016}
Galaxies in CL0016 show somewhat exceptional properties compared with other samples.
For example, as shown in Fig. \ref{fig:lf_env}, the field environment in CL0016
has a strange red LF, and the group environment has an increasing faint-end slope for red galaxies.
One may find that these LFs look different compared with other LFs.
The fraction of red galaxies is the highest in all environments as we see later.

There is a concern that this peculiarity may arise from errors in the
statistical subtraction of contamination.
We therefore check how the results for CL0016 can change by changing our
choice of field regions used in the subtraction.
We take either of the two field regions indicated by rectangles
in Fig. \ref{fig:cl0016_distrib} separately, rather than combining these two fields.
We then compare the resulting CMDs and LFs for CL0016 after the statistical subtraction.
This effectively corresponds to changing the field density by factor of $\sim2$.
Yet, we find no evidence that the results change strongly.
Therefore, we suggest that the peculiarity of the CL0016 cluster is intrinsic.
CL0016 is probably one of the oldest systems
in the Universe where red faint galaxies are already abundant.
Indeed, \citet{butcher84} illustrated the unusually low blue fraction of this
cluster for its redshift.
Our environments are defined on the basis of local and global (2Mpc) density.
It might be the case that even larger-scale environments (e.g., $\sim10$ Mpc)
can affect the galaxy properties
(\citealt{balogh04a}; but see also \citealt{blanton04}).
In the following, we regard CL0016 as an exceptional sample.

\subsection{Break Density}
\label{sec:break_density}
The break density corresponds to the outskirts of galaxy clusters
(one to two $R_{vir}$) and isolated groups.
This means that groups and clusters are dominated by red galaxies and supports the idea
that cluster-specific mechanisms have not played major roles in
transforming galaxy properties (\citealt{kodama01a,balogh04a,balogh04b,depropris04,tanaka04},
but see also \citealt{fujita04a,fujita04b}).
Rather, mechanisms such as low-velocity galaxy-galaxy interactions (e.g., \citealt{mihos96})
and strangulation \citep{larson80} remain as strong candidates of the driver
of the environmental dependence.

The fact that faint galaxies ($M_V>M^*_V+1$) show the clear break suggests
that physical mechanisms actually work on faint galaxies in high-density environments.
Then, why do not bright galaxies ($M_V<M^*_V+1$) show any prominent break?
Of course, bright galaxies are generally red and a break in their colours,
if any, would be difficult to see.
But, at least in the SDSS plot in Figure \ref{fig:colour_density}, this is not the case.
No break is seen even in the 25-th percentile of bright galaxies.
This implies that the evolutionary path of bright galaxies is different from that of
faint galaxies in such a way that the evolution of faint galaxies is strongly related to
groups and clusters, while that of bright galaxies is not strongly related \citep{tanaka04}.

Environmental dependence of galaxy properties is determined by
{\it a priori} effects (initial conditions) and {\it a posteriori} effects (environmental effects).
Recent near-IR studies such as {\it K20 Survey}
\citep{cimatti02a,cimatti02b,cimatti02c,daddi02,fontana04},
{\it FIRES} \citep{franx03,vandokkum03,rudnick03,schreiber04} and part of {\it GOODS}
\citep{giavalisco04,daddi04,moustakas04,somerville04} revealed
the existence of massive red galaxies at $z>1$.
Although a significant fraction of such red galaxies are dusty starbursts
(e.g., \citealt{miyazaki03}), massive non-star-forming galaxies,
which are presumably precursors of present-day ellipticals, do exist.
Since their colours match with passive evolution,
their properties are expected to be largely determined by {\it a priori} effects
and subsequent environmental effects are not very important.
That is, their star formation rate is already low before environmental mechanisms play a role.
These massive galaxies would evolve to bright galaxies in our definition,
and would explain the trends observed in \S \ref{sec:colour-density}.
On the other hand, the deficit of red faint galaxies is a function
of environment, in a way that red faint galaxies preferentially populate
in high-density environments.
Therefore, we suggest that properties of bright galaxies are largely determined by
{\it a priori} effects, while those of faint galaxies are largely determined
by {\it a posteriori} effects.

\subsection{The Build-up of the CMR}

In \S \ref{sec:colour-magnitude}, we observed the build-up of the CMR.
To further quantify the build-up, we base our analysis on stellar masses of galaxies.
If galaxies stop their star formation, they will be fainter
($\Delta M_V\sim1$) while keeping their stellar masses nearly unchanged.
Since our stellar mass estimates are not very accurate,
we cannot examine a detailed shape of a stellar mass function.
Instead, we investigate the giant-to-dwarf number ratio ($g/d$).
Giants and dwarves are defined as those having $\log_{10}(M_*/M_\odot)>10.6$ and
$9.7<\log_{10}(M_*/M_\odot)<10.6$, respectively.
Note that $\log_{10}(M_*/M_\odot)=10.6$ corresponds to $M_V=-20.5,\ -20.3$, and $-19.8$
for red galaxies at $z=0.83,\ 0.55$, and 0, respectively.

Since our estimates of stellar masses are primarily based on the the rest-frame
$V$-band magnitudes, they are affected by recent/on-going star formation
activities. In fact, as shown by the iso-mass curve in
Fig.~\ref{fig:cm_rest}, the $V$-band magnitude can change by 1.5 mag.
for the same stellar mass between passively evolving galaxies and the
constantly star forming galaxies.
This , which is then translated to the variation in
mass-to-light ratio by a factor of $\sim4$.
This can be viewed as a solid upper limit of the uncertainties in $M_*$ in
a relative sense,
since we correct for such variation in mass-to-light ratio by applying
SED fitting when deriving stellar masses
(note that absolute stellar masses depend on other factors
such as the stellar initial mass function).
Moreover, since we mainly discuss the red galaxies, the actual variation
in mass-to-light ratio must be much smaller (less than a factor of 2).
In what follows, we focus on the red galaxies only.
We separately discuss field, group, and cluster environments.

{\bf Field Environment:}
Figure \ref{fig:gd_evolution} shows $g/d$ in three different environments
and at three different redshifts.
The parameter $g/d$ clearly depends on both environment and time.
A general trend is that $g/d$ is the largest in the field environment
at any redshifts, suggesting that red faint galaxies is relatively rare
in the field.
This is consistent with our finding in Fig. \ref{fig:cm_rest} that
the faint-end of the field CMR is not clear.
The field $g/d$ ratio is the largest in SDSS.
This is likely to be driven by the build-up of the bright-end of the field CMR
(i.e., the fraction of giants increases).

{\bf Group Environment:}
The $g/d$ ratio in groups is the largest in RXJ0153 and it decreases
at low redshifts.
This decrease in $g/d$ may reflect the build-up of the group CMR at the faint-end:
the fraction of faint red galaxies increases.

{\bf Cluster Environment:}
The cluster $g/d$ also shows the decrease at low redshifts.
Although we could not see prominent build-up of the cluster CMR in \S \ref{sec:colour-magnitude},
the $g/d$ evolution may suggest the faint-end  of the cluster CMR is still under construction
even at $z=0.83$.

It seems that the bright-end of the CMR appears first,
and the faint-end is filled up later on.
A likely scenario for this build-up is that blue galaxies stop
their star formation and fade to settle down onto the CMR,
and the truncation of star formation starts from bright (massive) galaxies.
This scenario involves suppression of star formation activity in blue galaxies.
How do blue galaxies in groups and field environments
fade and settle exactly on the CMR as that of cluster galaxies?
We recall that we found no evidence for environmental dependence of the slope of the CMR.
If the CMR is a product of the mass-metallicity relation \citep{kodama97},
then blue galaxies have to 'know' the metallicities of red galaxies they settle onto.
Since blue galaxies in the field and group environments should follow
quite different star formation histories from those of red galaxies in clusters,
one may expect that their metallicities are quite different.
This is not the case, however.
An important point here is that we are considering relatively
massive ($\log_{10}M_*>9.7$) galaxies.
The cosmic star formation rate declines at $z<1$ (e.g., \citealt{madau96,madau98,fujita03}),
and therefore their major episode of star formation took place at higher redshifts.
Subsequent star formation does not strongly enrich metals in galaxies, and
metals locked in their stars do not grow significantly after the major star formation
\citep{tinsley80}.
Therefore, once galaxies mature, the epoch when they stop their star formation
is not a major concern from a chemical point of view.
Whenever they stop their star formation, they will settle onto the CMR.

To further quantify the build-up of the CMR, we plot in
Figure \ref{fig:red_fraction} the fraction of red galaxies
as functions of environment, stellar mass, and time.
As described in \S 3.3, we cannot deny the possibility that we miss a fraction
of blue galaxies in RXJ0153 and CL0016.
Therefore, the real red fractions in RXJ0153 and CL0016 would be smaller
than those shown in the plot.
We discuss the red fraction in the field, group, and cluster environments separately as follows.

{\bf Field Environment:}
The massive galaxies show an increase in the red fraction,
which is consistent with the build-up of the bright-end of the field CMR.
On the other hand, the least massive galaxies show a decrease.
However, this may be because we tend to miss a fraction of
blue galaxies in RXJ0153 and CL0016 due to our photo-z selection.

{\bf Group Environment:}
Due to the relatively large errors, the red fraction of group galaxies
is consistent with being unchanged with redshift.

{\bf Cluster Environment:}
Within the errors, red fraction is consistent with being almost constant over
the time under study.  This is in agreement with the trend that no significant
CMR build-up is seen in clusters within the redshift range explored.
There is an hint of evolution, however, at the least massive bin, where red
fraction may increase from $z=0.83$ to lower redshifts.

The overall trend is that the red fraction is the lowest in the field environment
and higher in groups and clusters.
But, in any environments and at any redshifts, the red fraction is higher
for more massive galaxies.
That is, the red fraction is higher in denser environment and for more
massive galaxies.

\subsection{Implications for Galaxy Evolution}

\begin{figure}
\begin{center}
\leavevmode
\epsfxsize 1.00\hsize \epsfbox{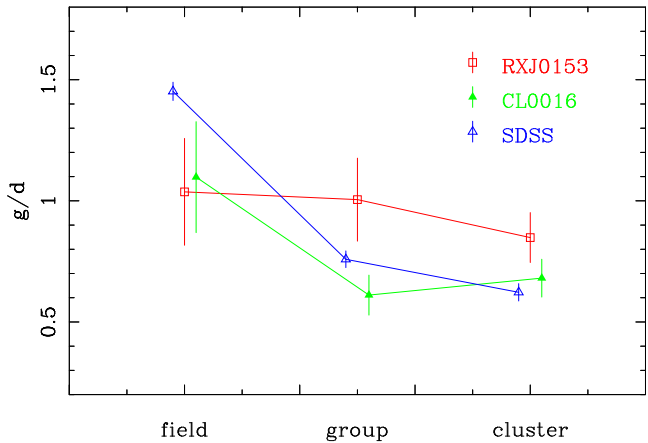}
\end{center}
\caption{
The giant-to-dwarf ratio of red galaxies plotted against environments.
The meanings of the lines are shown in the figure.
The errors are based on the Poisson statistics.
}
\label{fig:gd_evolution}
\end{figure}

\begin{figure}
\begin{center}
\leavevmode
\epsfxsize 1.00\hsize \epsfbox{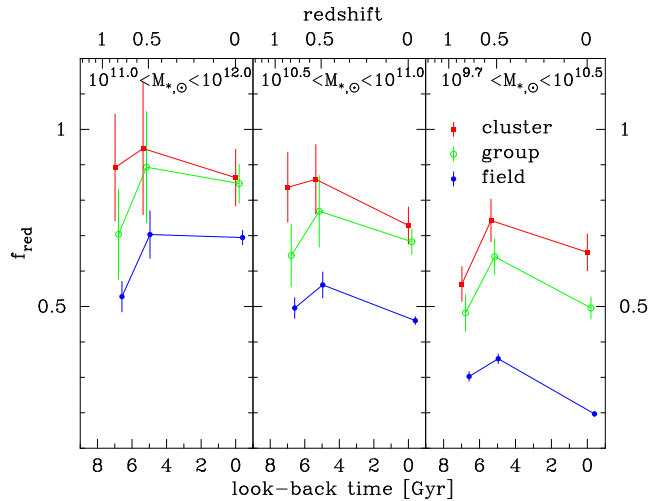}
\end{center}
\caption{
Fraction of red galaxies plotted against the look-back time
(the corresponding redshift is shown in the top ticks)
for the three stellar mass bins.
The meanings of the lines are shown in the figure.
Each point is shifted horizontally by a small amount for clarity.
The error bars show the Poisson errors.
}
\label{fig:red_fraction}
\end{figure}

From Figs. \ref{fig:gd_evolution} and \ref{fig:red_fraction}, we consider that
an evolutionary stage of the CMR build-up is different in different environments.
The bright-end of the cluster CMR is already built by $z=0.83$,
and only the faint-end shows a possible evolution at $z<0.83$.
On the contrary, in the field regions, the bright-end of the CMR
is being vigorously built, while the build-up at the faint-end has not started yet.

A possible interpretation of these results is that strong evolution has occurred
since $z\sim1$ in a 'down-sizing' way.
It was \citet{cowie96} who first pointed out the down-sizing galaxy evolution.
At high redshifts, massive galaxies actively form stars.
At lower redshifts, massive galaxies show less active star formation and
the main population of active star formation moves to less massive galaxies.
The bright-end of the CMR is formed first as massive galaxies stop their
star formation, and the build-up proceeds to the faint-end as less massive galaxies
stop their star formation.
Our results suggest that the evolutionary stage of this down-sizing depends on environment.
The evolution of massive cluster galaxies is almost completed by $z=0.83$.
In the field environment, the evolution of massive galaxies is strong,
while that of less massive galaxies is found to be weak.
Therefore, it seems that the main population that shows strong evolution
is shifted to higher mass galaxies in lower density environments.

One of the drivers of the environmental dependence of the down-sizing could be
initial conditions of galaxy formation ({\it a priori} effects).
Galaxies are formed earlier in higher density peaks of
the initial density fluctuation of the Universe.
Thus, galaxies in clusters are naturally at an advanced stage of galaxy evolution
compared with those in lower-density environments.
This may explain the environmental dependence of the build-up of the CMR.
We consider however that the environmental dependence of the down-sizing
effect is not solely caused by the {\it a priori} effects,
but environmental (or {\it a posteriori}) effects should contribute
significantly.
Effects that suppress star formation activities are strong in high-density environments
(\S \ref{sec:colour-density}), and they accelerate the build-up of the CMR.

We observed the build-up of the CMR at the faint-end in group environments.
This should mean either that faint galaxies are still actively forming stars,
or that less massive galaxies are not fully formed yet.
We cannot, however, discriminate these possibilities with the data in hand.
To do this, we need to estimate stellar masses of faint blue galaxies accurately.
Further discussion on the driver of the CMR build-up awaits
more accurate stellar mass estimates by near-infrared data.

Detailed observations of early-type galaxies suggest that the typical luminosity-weighted
age of field early-type galaxies is younger than cluster galaxies
(e.g., \citealt{kuntschner02,gebhardt03}).
\citet{thomas04} examined nearby early-type galaxies and suggested that
formation of early-type galaxies is the earliest in high density environments
and delayed by $\sim2$ Gyr in low density environments, and
formation of massive galaxies predates that of less massive galaxies.
Utilizing near-IR data, \citet{feulner04} and \citet{juneau04} reported that
the epoch of major star formation took place at higher redshifts for more massive galaxies,
and star formation is more extended in time for less massive galaxies.
All these studies lend support to the down-sizing picture.

We close our discussion by noting some caveats on our results.
Our results are based only on two high-$z$ clusters,
which may not be a typical cluster at each redshift (particularly CL0016).
To reach a firm conclusion, we need to observe more clusters at various redshifts.
Also, errors in the photometric redshifts and stellar mass estimates remain as a concern.
Near-infrared (e.g., $K$-band) data are required to improve them.
Spectroscopic data are clearly needed to further address the effectiveness of
the photometric redshifts and assess errors in the statistical contamination subtraction.
We hope to overcome these uncertainties in our future paper.

%
%
\section{Summary and Conclusions}
\label{sec:conclusion}
We began this paper by introducing the three axes on which galaxy properties
strongly depend, namely, environment, stellar mass, and time.
We found that the star formation activity in galaxies is indeed dependent on
all of these three axes and galaxies follow complicated evolutionary paths.

We conducted multi-band imaging of two high-$z$ clusters, RXJ0153 at $z=0.83$ and
CL0016 at $z=0.55$, with Suprime-Cam on Subaru.
These Subaru data were combined with the SDSS data ($z=0$),
and we carried out statistical analyses of galaxy properties
as functions of environment, mass, and time.
We examined the colour-density relations, colour-magnitude diagrams (CMDs), and
luminosity functions (LFs) of galaxies.

First, we applied our photometric redshift technique to RXJ0153 and CL0016 fields
to largely eliminate fore-/background contamination and discovered large-scale structures
surrounding the main clusters.
Details are described in \citet{kodama05}.

Then we examined the relationship between galaxy colours and environments.
It was found that galaxy colours abruptly change at the break density
at any redshifts considered in this paper.
Faint galaxies ($M_V>M_V^*+1$) show a prominent break, while
bright galaxies ($M_V<M_V^*+1$) do not show such a strong break.
Based on the break density, we defined three environments:
field, group, and cluster.

We examined CMDs in the three environments.
The highlight of the CMD analysis was to show how the CMR is built up as
functions of time, environment and mass.
There is no clear CMR in the field regions in RXJ0153,
while a clear CMR can be seen in the same environment in CL0016 and SDSS,
particularly at the bright-end.
In groups of RXJ0153, a clear CMR is visible only at the bright-end ($M_V<-20$).
The relation is found to extend down to our magnitude limit in groups of CL0016 and SDSS.
Clusters have clear CMRs down to our magnitude limits
at all redshifts considered here, with a possible remaining
activity at the faint-end.
These trends are quantified by the scatter of the CMRs.

The build-up of the CMR was quantified by the giant-to-dwarf ratio
($g/d$) and the red fraction.
From $z=0.83$ to $z=0$, the field $g/d$ increases and the group $g/d$ decreases,
suggesting the CMR build-up at the bright-end in the field and at the faint-end in the group.
The cluster $g/d$ also decreases.  This may imply that the faint-end of the cluster CMR
was still under construction at high redshifts.
We found that the red fraction is higher in denser environment and for more massive galaxies.

As a possible interpretation of these results, we suggested that galaxies evolve
in the down-sizing way.
That is, the main populations that host active star formation is shifted
from massive galaxies to less massive galaxies with time.
It is likely that an evolutionary stage of the down-sizing depends on environment.
Cluster galaxies evolve most rapidly, and group and field galaxies follow.
All in all, it seems that galaxy evolution proceeds from massive galaxies to
less massive galaxies, and from high density environments to low density environments.
In order to confirm this trend, however, further studies are obviously needed.
We need to increase the cluster sample and also to perform follow-up
observations such as NIR imaging and spectroscopic surveys.

%
%
\section*{Acknowledgments}
M.T. acknowledges support from the Japan Society for Promotion of Science (JSPS)
through JSPS research fellowships for Young Scientists.
We thank Fumiaki Nakata and Masafumi Yagi for their help during the data reduction.
We thank the anonymous referee for the careful reading of the paper and invaluable
comments which improved the paper significantly.
This work was financially supported in part by a Grant-in-Aid for the
Scientific Research (No.\, 15740126, 16540223) by the Japanese Ministry of Education,
Culture, Sports and Science.
This study is based on data collected at Subaru Telescope, which is operated by
the National Astronomical Observatory of Japan. 

Funding for the creation and distribution of the SDSS Archive has been provided
by the Alfred P. Sloan Foundation, the Participating Institutions,
the National Aeronautics and Space Administration, the National Science Foundation,
the U.S. Department of Energy, the Japanese Monbukagakusho, and the Max Planck Society.
The SDSS Web site is http://www.sdss.org/.
The SDSS is managed by the Astrophysical Research Consortium (ARC)
for the Participating Institutions. The Participating Institutions
are The University of Chicago, Fermilab, the Institute for Advanced Study,
the Japan Participation Group, The Johns Hopkins University, the Korean Scientist Group,
Los Alamos National Laboratory, the Max-Planck-Institute for Astronomy (MPIA),
the Max-Planck-Institute for Astrophysics (MPA), New Mexico State University,
University of Pittsburgh, Princeton University, the United States Naval Observatory,
and the University of Washington.

%
%

%
%
\appendix

\section{Details of the Volume Correction}
\label{app:vmax}
In a flux-limited sample, intrinsically faint galaxies are observed only at
low redshifts, while intrinsically bright galaxies can be observed at higher redshifts.
For example, in our SDSS sample, we can observe galaxies brighter than
$M_{V,\ vega}<-19.5$ at all redshifts, while $M_{V,\ vega}<-18$ can be observed
only at $z<0.033$.  We therefore need to correct for this incompleteness when 
we discuss galaxy properties.
A simple way to do this is to give heavier statistical weights to
intrinsically fainter galaxies according to their redshifts.
A flux-limited sample can mimic a volume-limited one in this way.
A statistical weight can be computed as

\begin{equation}
w_i=V_{survey}/V_{i, max},
\end{equation}

\noindent
where $w_i$ is a statistical weight of $i$-th galaxy, $V_{survey}$
is the volume contained within our redshift range ($0.005<z<0.065$),
and $V_{i,max}$ is the maximum volume over which the $i$-th galaxy could be
observed (e.g., in case of $M_V=-18.0$, $V_{max}$ is the volume
contained within $0.005<z<0.033$).
We assume that a LF does not evolve within the surveyed redshift range.

We can test effectiveness of this volume correction by simulating
galaxy distribution with a given LF.  For simplicity, we adopt a flat LF
such that galaxies are uniformly distributed within a magnitude range of $-21<M<-15$.
We examine a set of galaxy distribution here.

\begin{figure}
\begin{center}
\leavevmode
\epsfxsize 1.0\hsize \epsfbox{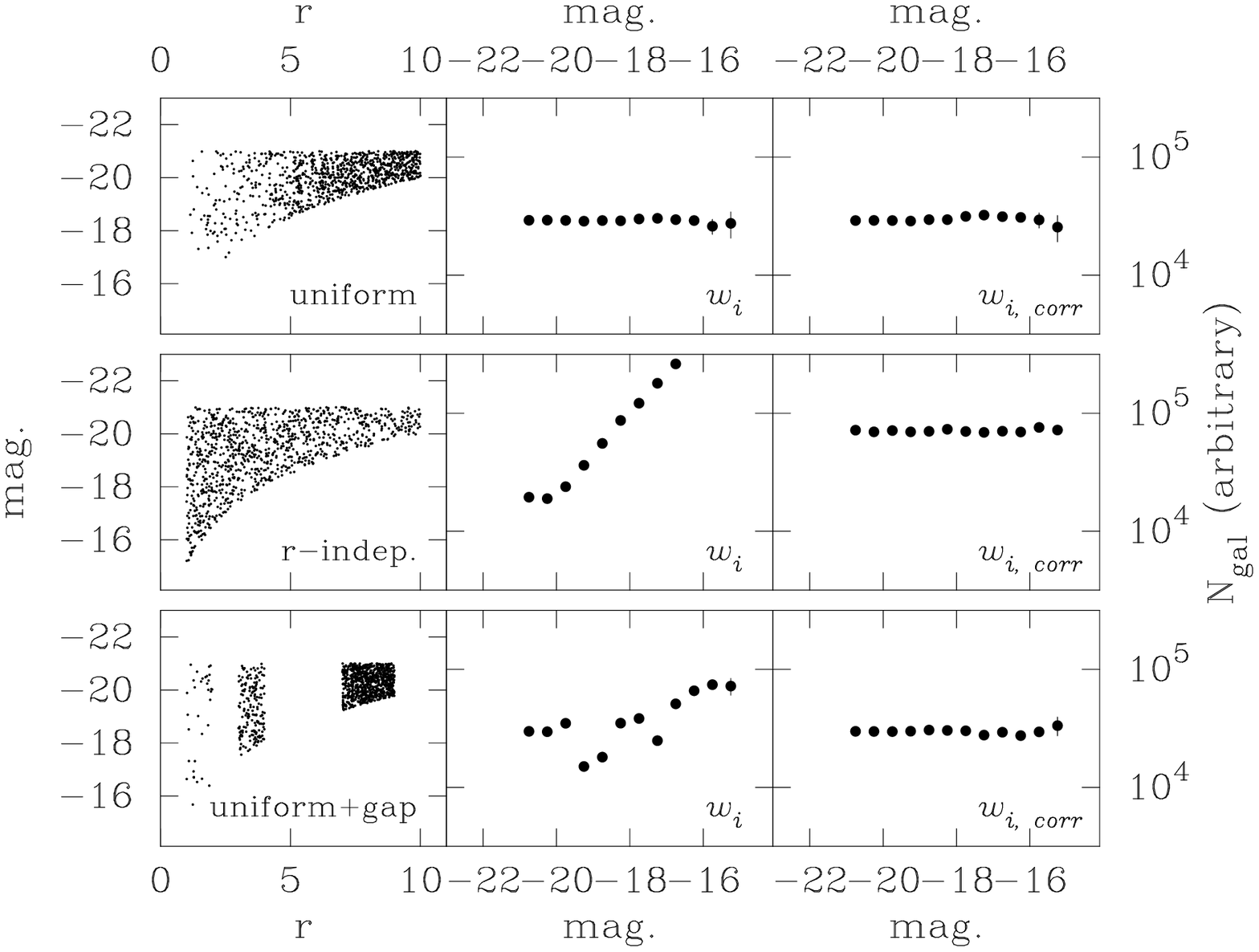}
\end{center}
\caption{
The volume corrected LFs of three types of galaxy distribution.
>From top to bottom, the panels show uniform distribution ($N_{gal}\propto r2$),
r-independent distribution ($N_{gal}$ is constant at a given $r$), and
uniform distribution with redshift gaps.
The left panels show galaxy distribution as a function of a distance from us
(unit is arbitrary).
The middle panels present reconstructed LFs based on the $w_i$ weighting.
The right panels show reconstructed LFs based on the $w_{i,corr}$ weighting.
See text for details.
}
\label{fig:vmax_sim}
\end{figure}

In the top panels of Figure \ref{fig:vmax_sim}, we show a model of
uniform galaxy distribution in a given volume, and the middle
of the top panels shows a volume corrected LF.
As seen in the panel, a LF is reconstructed well.
In the middle panels, we set non-uniform galaxy distribution in a given volume.
If galaxies are uniformly distributed in a given volume, we expect that
the number of galaxies in a given radial shell scales as  $N_{gal}\propto r2$
(this is the case for the top panels).
But, here we adopt the distribution such that $N_{gal}$ is independent of $r$.
In this case, galaxies are not distributed uniformely in a volume, and
a volume corrected LF does not reconstruct the parent LF.
This is expected -- the volume correction assumes the uniform
galaxy distribution over a survey volume.
The bottom panels show another example, where radial galaxy distribution
has 'gaps'.  The volume correction again fails to reproduce the parent LF.

Therefore, we take into account variations in radial galaxy distribution
such as large-scale structures when we perform the volume correction.
In our sample, galaxies having $M_V<-19.5$ can be observed regardless of redshift
(i.e., volume-limited).
We use these galaxies to correct for the radial variation.
A number of galaxies with $M_V<-19.5$ should scale as  $\propto r2$
if they are distributed uniformly in the Universe.
We define radial distribution of galaxies having $M_V<-19.5$ divided by $r2$
as $f(r)$, and radially averaged $f(r)$ as $f_{ave}$.
A statistical weight becomes:

\begin{equation}
\label{eq:volume_corr}
w_{i,corr}=\left( V_{survey}/V_{max}\right) \times \left[ f_{ave}/f(r)\right].
\end{equation}

\noindent
Note that, in the first term, we do not include the volume
where no galaxies are observed.
For example, we do not calculate the volume contained in the 'gaps' in
the bottom panels in Fig. \ref{fig:vmax_sim} because
no statistical correction can be made where there are no galaxies.
In the latter term, we make radial bins to compute $f(r)$ and $f_{ave}$.
We confirmed that our conclusions are not strongly affected by uncertainties
arising from the binning.
The right panels of Fig. \ref{fig:vmax_sim} shows volume corrected LFs
using this equation.  The parent LFs is reconstructed well.
Note that Eq. \ref{eq:volume_corr} does not reproduce an absolute number of galaxies.

As a final check of this correction, we use real galaxy distribution
in our sample and artificially assign their magnitudes based on a flat LF.
We confirm that the parent LF is recovered well in all the field, group,
and cluster samples by using Eq. \ref{eq:volume_corr}.
It turns out that the correction for the non-uniformity is particularly important
for group and cluster galaxies whose distribution is strongly related to
large-scale structures.
Actually, cluster galaxies have 'gaps' in their radial distribution.
To sum up, a flux-limited sample can be used to discuss galaxy properties
by applying statistical weights of Eq. \ref{eq:volume_corr}.

\section{Statistical Contamination Subtraction}
\label{app:field_sub}

\begin{figure}
\begin{center}
\leavevmode
\epsfxsize 1.0\hsize \epsfbox{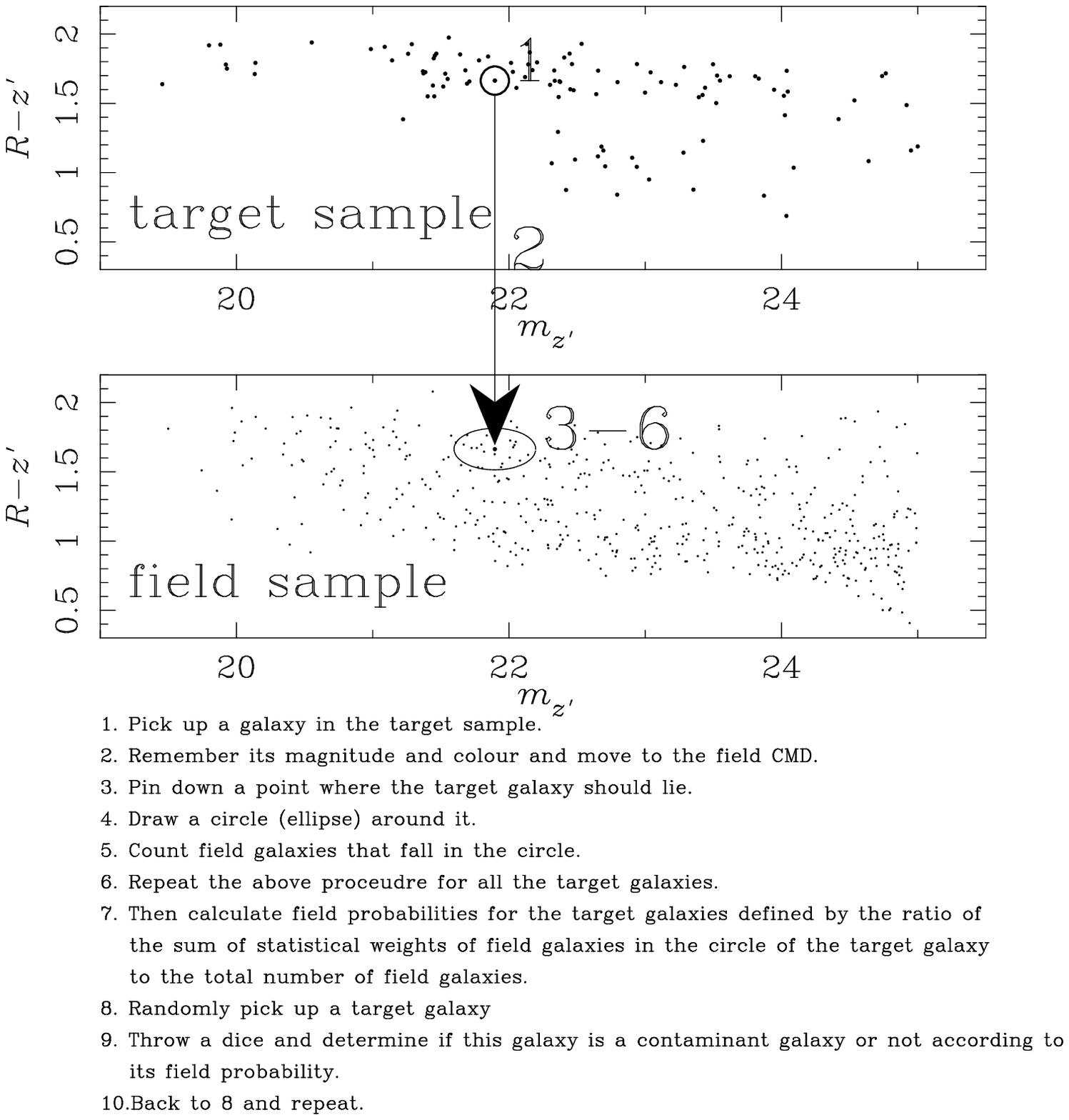}
\end{center}
\caption{
The contamination subtraction procedure.
}
\label{fig:field_sub}
\end{figure}

This appendix describes the statistical contamination subtraction procedure in detail.
See Figure \ref{fig:field_sub} for help.
We use the control field sample defined in \S\ref{sec:field_sub} to
subtract the contamination.
For simplicity, in this appendix, we refer to the control field sample
as field sample.  The field sample examined in the main text should be
considered as a 'target' sample here.

Let us pick up a galaxy in a target CMD.
We move on to a CMD of field galaxies,
and pin down a point where our target galaxy should lie (i.e.,
magnitude and colour of the galaxy we picked up).
Then we draw a circle (strictly speaking, ellipse) around the
point on the field CMD, and count field galaxies
that fall in the circle.
The above procedure is performed for all target galaxies, and then we know
how many times a field galaxy is counted by the target galaxies.
We define a statistical weight for a field galaxy as an inverse
of its count.
For example, if a field galaxy is counted three times (i.e.,
three target galaxies share the field galaxy),
the statistical weight of the field galaxy is 1/3.
Then we define a probability for a target galaxy to be a field
galaxy as the ratio of the sum of statistical weights of field galaxies
in the circle of the target galaxy to the total number of field galaxies.
We randomly pick up a target galaxy, throw a dice, and determine
if the target galaxy is a field galaxy or not according to its
field probability.
We repeat this procedure until we subtract sufficient number of
galaxies.
In short, this procedure can be considered as a generalized
form of the grid-based method \citep{kodama01b,pimbblet02}.
It should be noted that since we make no grids, we are free from
an uncertainty on how we make the grids.
Also note that we are free from the 'negative galaxy' problem which
the grid-based method faces.

There are two parameters in our method, namely, the radius and the shape
of the aperture within which we count field galaxies.
The aperture we actually use is shown in Figure \ref{fig:field_sub}.
It should be noted that our results do not depend on
the size and shape of the aperture in reasonable ranges.

How many galaxies should we subtract?
When we have galaxies that occupy surface area of ${\mathcal A}\rm\ arcmin2$,
we expect the contamination to be ${\rm \Sigma_{field}} \mathcal{A}$ galaxies,
where ${\rm \Sigma_{field}}$ is an average surface density of field galaxies.
Since we define environment by galaxy density,
it is not very straightforward to estimate a surface area occupied by density-selected galaxies.
We infer a surface area by a simple approximation.
Because an inverse of a galaxy density means an average surface area occupied by a galaxy
(i.e., unit of ${\rm arcmin^{2}}\ galaxies^{-1}$), a sum of an inverse of density
gives a surface area:

\begin{equation}
\label{eq:density2area}
{\mathcal A}_\Sigma=\sum_{i=1}^{N}\frac{1}{\Sigma_{local, i}},
\end{equation}

\noindent
where $\Sigma_{local}$ is local density and $i$ runs through 1st to $N$-th galaxy we select.
We show in Appendix \ref{app:conversion} that this estimate is statistically robust.
A Poisson error is then added in ${\rm \Sigma_{field}} {\mathcal A}_\Sigma$
in each Monte-Carlo realization of the contamination subtraction.

\begin{figure}
\begin{center}
\leavevmode
\epsfxsize 1.0\hsize \epsfbox{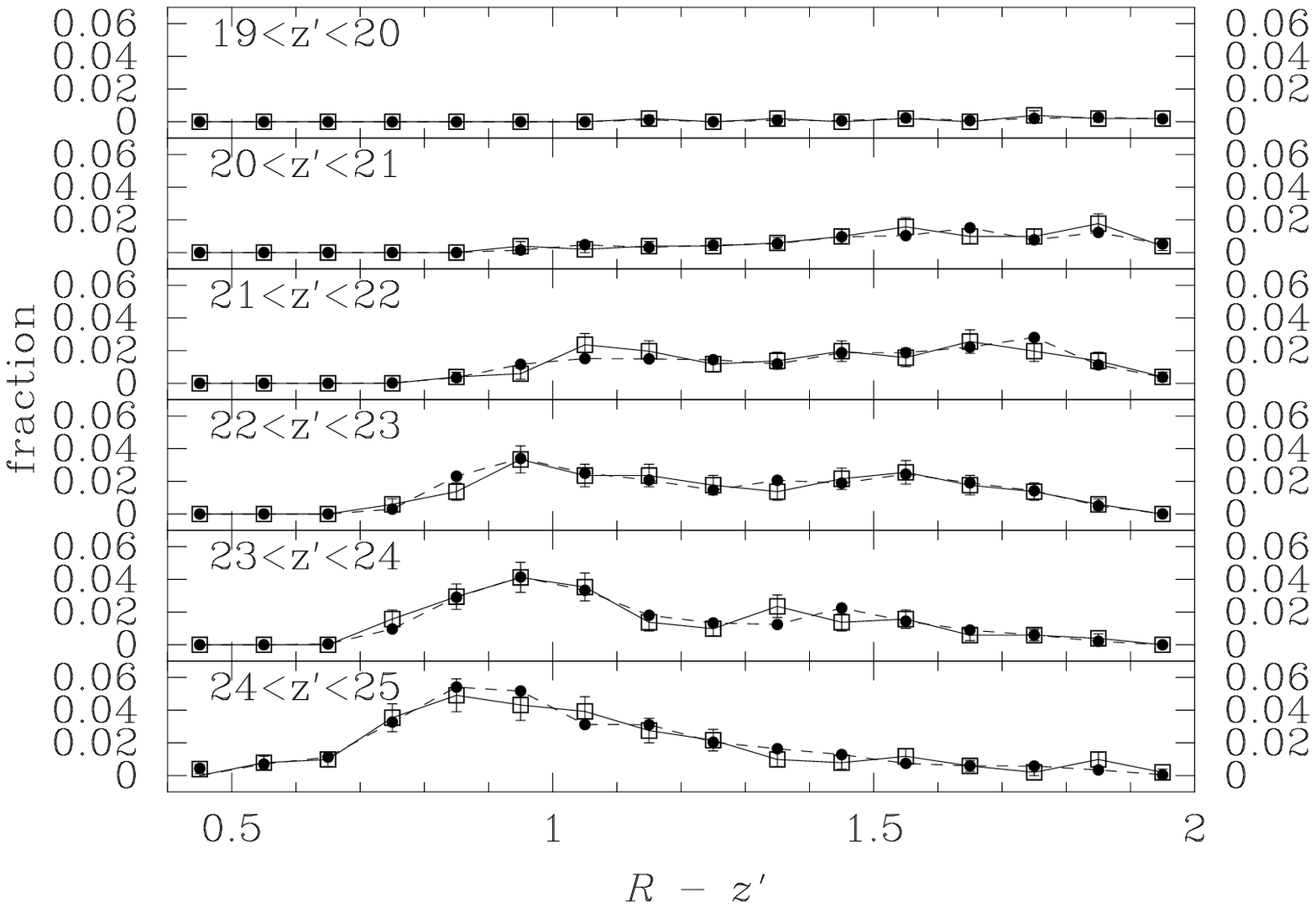}
\end{center}
\caption{
Field galaxies in RXJ0153 are divided into 6 magnitude bins (6 panels from top to bottom),
and fractions of field galaxies as a function of $R-z'$ colour are plotted.
The open squares show distribution of field galaxies.
The solid circles show distribution of 'subtracted' target galaxies.
}
\label{fig:consistency2}
\end{figure}

We check if our statistical contamination subtraction technique
properly subtracts field galaxies.
We examine if subtracted target galaxies reproduce the CMD of field galaxies.
Figure \ref{fig:consistency2} plots distribution of subtracted target galaxies on a CMD
projected onto 1 dimensional space for clarity.
It can be seen that the distribution of subtracted galaxies closely follows that of
field galaxies.
The very small difference between them is due to the fact that 
we apply a smoothing of the galaxy distribution on the field CMD
when we estimate field probabilities (the size and shape of the aperture determines the
smoothing scale).
Despite the small difference we see a fairly good correspondence, and 
we therefore conclude that our subtraction method properly removes the field contamination.


\section{Conversion from Density to Surface Area}
\label{app:conversion}

In Appendix \ref{app:field_sub}, we have converted from local density to
a surface area that is occupied by density-selected galaxies.
We examine validity of this approximation in this section.
We construct a simple toy model of galaxy cluster and investigate differences between
the real (metric) area and the area estimated from density.

Our toy model consists of uniform field galaxies and cluster galaxies whose distribution follows
the King model \citep{king62}.
We randomly generate galaxies and estimate density for each galaxy by
the nearest-neighbor method used in this paper.
Galaxies within a fixed aperture around the cluster is selected.
Then we compare the area of the aperture (${\mathcal A}_{aper}$) with the area estimated from densities
of galaxies within the aperture (${\mathcal A}_{\Sigma}$, see eq. \ref{eq:density2area}).
We calculate ${\mathcal A}_{\Sigma}/{\mathcal A}_{aper}$ in each realization and repeat 1000 times.
There are two parameters in the model: the radius of the aperture,
and the fraction of cluster galaxies to field galaxies in the aperture.

\begin{figure}
\begin{center}
\leavevmode
\epsfxsize 1.0\hsize \epsfbox{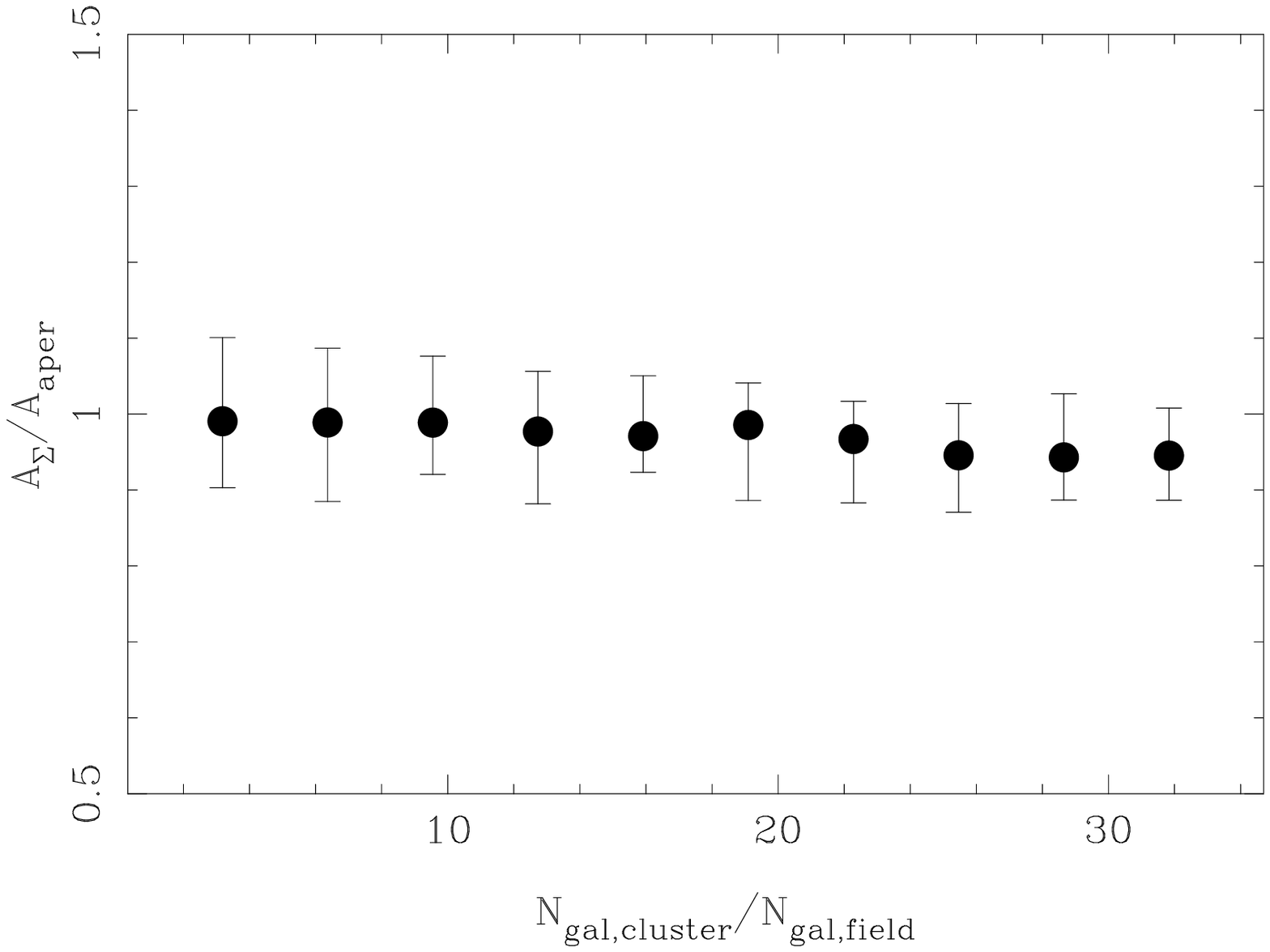}
\end{center}
\caption{
${\mathcal A}_\Sigma/{\mathcal A}_{aper}$ plotted against the number ratio of cluster galaxies to
field galaxies in the aperture. The points mean the median of the distribution,
and the error bars represent the quartiles of the distribution.
}
\label{fig:density_area}
\end{figure}

As a fiducial value, the aperture radius is set to $3r_c$
($r_c$ means the core radius of the King model).
Figure \ref{fig:density_area} plots ${\mathcal A}_{\Sigma}/{\mathcal A}_{aper}$
against the number ratio of cluster galaxies to field galaxies in the aperture.
The result is encouraging -- ${\mathcal A}_{\Sigma}/{\mathcal A}_{aper}$ is almost unity.
In the density calculation, the distance to n-th nearest galaxy is used
($n=5$ and 10 is adopted in the main text).
We run the simulation using $n=5,\ 10,\ 20$, and find that there is a weak trend
that the scatter in ${\mathcal A}_{\Sigma}/{\mathcal A}_{aper}$ decreases with increasing $n$,
but the median is always close to unity.
Our choice of the aperture radius seems to have a rather strong effect.
The scatter in ${\mathcal A}_{\Sigma}/{\mathcal A}_{aper}$ increases as we adopt smaller aperture (e.g., $2r_c$). 
On the other hand, if we adopt larger aperture, say $5r_c$, the scatter reduces to almost half.
However, in any case, the median ${\mathcal A}_{\Sigma}/{\mathcal A}_{aper}$ is always close to unity.
Therefore, we conclude that ${\mathcal A}_{\Sigma}$ is a good measure in a statistical sense.

\section{Global Density}
\label{app:global_density}

Our aim in this appendix is to address how effective and quantitative
our group/cluster separations are.
For simplicity, groups and clusters are referred to as systems in this Appendix.

We run the friends-of-friends algorithm (FOFA; \citealt{huchra82}) in our SDSS sample to
find galaxy groups and clusters.
The FOFA is a famous group/cluster finding algorithm and its statistical
properties are well-known \citep{moore93,frederic95,ramella97,diaferio99,mercian02}.
The basic algorithm of the FOFA is to find a set of galaxies connected within
certain linking lengths from each other.
There are two linking lengths, namely, an angular separation ($D_0$) and
a line-of-sight velocity difference ($V_0$).
Since galaxy systems are elongated along the line-of-sight in the redshift space
('finger of god' effect), we need to handle $D_0$ and $V_0$ separately.

We run the FOFA in a volume-limited sample defined by
$0.010<z<0.065$ and $M_g<-19.3$ so that we do not have to apply any
scaling in the linking lengths.
The FOFA parameters are set to $D_0$=700kpc and $V_0=400\kms$.
These values are slightly different from those adopted in \citet{tanaka04}.
Our conclusions are not, however, strongly dependent on the choice of the parameters.
There is one more parameter, $N_{min}$, which determines the minimum size of galaxy systems.
If a system has members less than $N_{min}$, the system is not considered here.
We set $N_{min}=5$ as in \citet{tanaka04} since a significant fraction of $N<5$ groups are
expected to be spurious \citep{frederic95,ramella97,ramella02}.

\begin{figure}
\begin{center}
\leavevmode
\epsfxsize 1.0\hsize \epsfbox{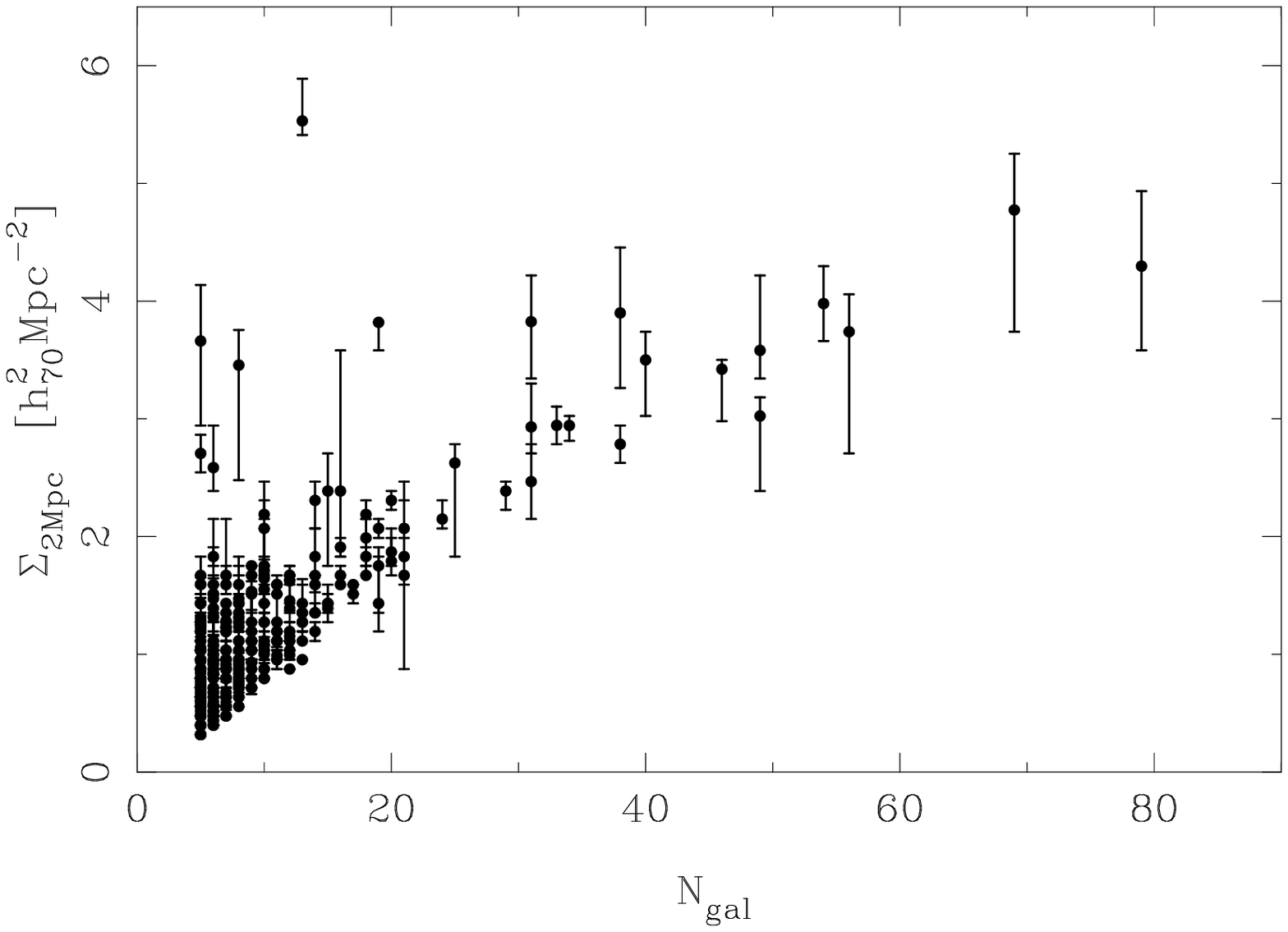}
\end{center}
\caption{
Global density plotted against $N_{gal}$ determined by the FOFA.
The dot and the error bars show the median and quartiles of global density distribution in each system.
}
\label{fig:ngal_vs_global}
\end{figure}

In Figure \ref{fig:ngal_vs_global} we present a correlation between
the number of member galaxies ($N_{gal}$) of the FOFA systems and
global density of the member galaxies.
The correlation is encouragingly good -- there is a positive correlation between
the two quantities, especially at $N_{gal}>20$.
$N_{gal}<20$ systems show only a weak correlation.
This is probably because that a typical extent of such poor systems is smaller than
2Mpc and a number fluctuation of field galaxies contribute to global density.
As for $N_{gal}>20$, cluster members dominate the 2Mpc aperture and hence
we have a good correlation there.
A few systems that have small $N_{gal}$ with high global density can be seen.
They are either parts of rich systems accidentally splitted by the FOFA
or real isolated systems close to nearby rich systems.
They comprise only a small fraction and the effect of such contamination is negligible.
In summary, global density is a powerful tool to separate groups from clusters quantitatively.

We measure velocity dispersions of detected systems using the biweight estimator \citep{beers90}.
By examining the relationship between $N_{gal}$ and a velocity dispersion of a system
($\sigma$), we find that the global density threshold used in the main text
for the SDSS, $\Sigma_{\rm global}=2.5$, roughly corresponds to systems having
$\sigma=300-400\kms$.
Therefore, our groups typically have $\sigma<300-400\kms$ and
clusters have $\sigma>300-400\kms$.

\end{document}